\documentclass[11pt,tightenlines,eqsecnum,floats,aps,amsmath,amssymb,nofootinbib,prd,shownopacs,floatfix]{revtex4}

\usepackage[caption=false]{subfig}
\usepackage{amsmath,amssymb,amsfonts,dcolumn,color,graphicx,graphics,latexsym}
\usepackage[mathscr]{eucal}
\usepackage[latin1]{inputenc}
\usepackage{enumerate}
\usepackage{accents}

\newcommand\munderbar[1]{%
  \underaccent{\bar}{#1}}

\newcommand{\be}{\begin{equation}}
\newcommand{\ee}{\end{equation}}
\newcommand{\ba}{\begin{eqnarray}}
\newcommand{\ea}{\end{eqnarray}}

\newcommand{\lp}{l_{\mathrm{Pl}}}

\newcommand{\bmult}{\nopagebreak[3]\begin{multline}}
\newcommand{\emult}{\end{multline}}

\def\d{{\rm d}}
\def\p{\partial}

\renewcommand{\inf}{\infty}
\begin{document}

\title{Emergence of product of constant curvature spaces in loop quantum cosmology}
\author{Naresh Dadhich$^{*,\dagger}$}
\author{Anton Joe$^\ddagger$}
\author{Parampreet Singh$^\ddagger$}
\affiliation{${}^*$Inter-University Center for Astronomy and Astrophysics, \\
Ganeshkhind, Pune 411 007, India \\
\,
\,
\,
\\
${}^\dagger$ Center for Theoretical Physics, Jamia Millia Islamia, Delhi, India \\
\,
\,
\,
\\
${}^\ddagger$ Department of Physics and Astronomy,
Louisiana State University, Baton Rouge, LA 70803, U.S.A.}

\begin{abstract}
The loop quantum dynamics of Kantowski-Sachs spacetime and the interior of higher genus black hole spacetimes with a cosmological
constant has some peculiar features not shared by various other spacetimes in loop quantum cosmology. As in the other cases,
though the quantum geometric effects resolve the physical singularity and result in a non-singular bounce, after the bounce
a spacetime with small spacetime curvature does not emerge  in either the subsequent backward or the forward evolution.
Rather, in the asymptotic limit the spacetime manifold is a product of two constant curvature spaces.
Interestingly, though the spacetime curvature of these
asymptotic spacetimes is very high, their effective metric  is a solution to the Einstein's field equations.
 Analysis of the components of the Ricci tensor shows that after the singularity resolution, the  Kantowski-Sachs spacetime
leads to an effective metric which can be interpreted as the `charged' Nariai, while the higher genus black hole
interior can similarly be interpreted as  anti Bertotti-Robinson spacetime
with a cosmological constant. These spacetimes are `charged'
in the sense that the energy momentum tensor that satisfies the Einstein's field equations is formally the same as
the one for the uniform electromagnetic field, albeit it has a purely quantum geometric origin.  The asymptotic spacetimes also have
an emergent cosmological constant which is different in magnitude,
and sometimes even its sign, from the cosmological constant in the Kantowski-Sachs and the interior of higher genus black
hole metrics. With a fine tuning of the latter cosmological constant, we show that `uncharged' Nariai, and  anti Bertotti-Robinson spacetimes
with a vanishing emergent cosmological constant can also be obtained.
\end{abstract}

\maketitle

\section{Introduction}

The central singularity in the black hole spacetimes is the boundary of  validity of the general relatvistic description in a
gravitational collapse. At this singularity, the curvature invariants diverge and the geodesic evolution breaks down. A fundamental issue
is to understand the way quantum gravitational effects lead to the singularity resolution, and what is the nature of the emergent spacetime beyond the
classical singularity.  In recent years, techniques of loop quantum gravity (LQG), a candidate theory for quantum gravity based on 
connection and triad variables,
 have been extensively applied to understand the fate of singularities in
various gravitational spacetimes which exhibit symmetries, such as in the cosmological setting in Loop Quantum Cosmology (LQC) (for a review see, Ref. \cite{as}), as well as for
the black hole spacetimes \cite{ab,mod,gp2,pullin,gambini,chiou}. In homogeneous isotropic and anisotropic spacetimes, quantum dynamics is 
obtained from the quantum Hamiltonian constraint which turns out to be a difference equation with discreteness determined by the quantum 
geometry in LQG. At small spacetime curvature, the quantum discreteness becomes negligible and Wheeler-DeWitt equation is recovered. 
Departures from general relativity manifest near the Planck scale, whereas in the infra-red limit, loop quantization of the above models 
leads to classical dynamics. A key result in LQC is the existence of a bounce at the Planck scale which is a direct result of the 
underlying discrete quantum geometry \cite{aps1,aps2,aps3}. Using an exactly solvable model \cite{slqc}, the consistent probability of 
bounce has been computed, which turns out to be unity \cite{craigsingh}. Extensive numerical simulations, so far carried out for isotropic and Bianchi-I spacetimes,
confirm that physical states obtained from solving the quantum Hamiltonian constraint remain peaked 
on the classical trajectory  for a long time before entering the Planck regime
where it bounces to another classical branch  with a small spacetime curvature \cite{aps3,apsv,bp,ap,ps12,num2,rad,num3,b1_madrid,num4}. 
The relative fluctuations of the physical states remain tightly constrained across the bounce \cite{cs2,kp2,cm1,cm2}, and the physics of 
this non-singular evolution
can be described using an
effective spacetime description which turns out to be an excellent approximation for a large class of states \cite{num2,num3,num4}.
The effective Hamiltonian faithfully captures the quantum geometric degrees of freedom and the quantum discreteness encoded in the holonomies of the connection in a
continuum spacetime. Thanks to the
availability of the effective Hamiltonian, quantum geometric effects in LQC have been extensively studied via the modifications to the classical Hamilton's equations
in general relativity (GR) \cite{as}.

In the context of the black holes, effective Hamiltonian techniques in LQC can again be employed to gain insights on the Planck scale 
physics in the interior spacetime.
In particular, the Schwarzschild black hole interior corresponds to the vacuum Kantowski-Sachs spacetime. Similarly, Schwarzschild de Sitter and Schwarzschild anti-de Sitter black hole
interiors can also be studied in minisuperspace setting using Kantowski-Sachs cosmology with a positive and a negative cosmological constant respectively. Additionally the Bianchi III LRS spacetime which is analogous to Kantowski-Sachs
spacetime but with a negative spatial curvature, turns out to be corresponding to the higher genus black hole interior.  Using symmetries 
of these spacetimes, the connection and triad variables simplify and a rigorous loop quantization can be performed which results in a 
quantum difference equation, and an effective spacetime description. Analysis of the expansion and shear scalars 
for the loop quantized Kantowski-Sachs spacetime reveal that there exists a unique quantization prescription which leads to their 
universally bounded behavior \cite{js1}.  Similar conclusions hold for the higher genus black hole interiors, as is discussed in this 
manuscript. 
This quantization prescription corresponds to the so called `improved dynamics' in LQC \cite{aps3}.   Using the corresponding effective 
Hamiltonian approach for this quantization prescription, singularity avoidance via a quantum bounce due to underlying loop quantum 
geometric effects in black hole interior spacetimes has been found \cite{bv,dwc,Brannlund}. 
These studies noted that the emergent spacetime is ``Nariai type" 
\cite{bv,Brannlund}. Further, these ``Nariai type" spacetimes were found to be stable under homogeneous perturbations in the case of vacuum 
\cite{bv1}. However, the detailed nature of these spacetimes and their relation if any with the known spacetimes in the classical theory 
was not found. An examination of these spacetimes, which is a goal of this manuscript, reveals many interesting features which so far 
remain undiscovered in LQC.

The spatial manifold of Kantowski-Sachs spacetime has an $\mathbb{R} \times \mathbb{S}^2$ topology whereas the higher genus 
black hole interior has the spatial topology of $\mathbb{R} \times \mathbb{H}^2$. Numerically solving the loop quantum 
dynamics one finds that on one side of the temporal evolution, in the asymptotic limit, the spacetime emergent after the bounce has the 
same spatial topology, but has a constant radius for the $\mathbb{S}^2$ ($\mathbb{H}^2$ in the case of higher genus black hole) part and 
an exponentially increasing $\mathbb{R}$ part. We thus obtain a
spacetime which is a product of two constant curvature spaces. Interestingly, though the emergent spacetime has a high spacetime 
curvature, yet it turns out to be a solution of the Einstein's field equations. In the analysis of  these spacetimes, the sign of the Ricci 
tensor components provide important insights. Here we recall that in the classical GR, properties of the sign of the Ricci tensor components 
have been used to establish dualities between (anti) Nariai and (anti) Bertotti Robinson spacetimes \cite{Dadhich}. Analysis of the 
components of the Ricci tensor 
reveals that  the emergent spacetime in the evolution of 
Kantowski-Sachs
spacetime with positive or negative cosmological constant is a `charged' Nariai spacetime, where as the emergent spacetime in the evolution of higher genus black hole interior with a negative cosmological constant  is actually an anti-Bertotti-Robinson spacetime with a cosmological constant. These emergent spacetimes are `charged' in the
sense that they are solutions of the classical Einstein's field equations with a stress energy tensor which formally corresponds to 
the uniform electromagnetic field. In addition, these spacetimes in the same asymptotic limit after the bounce also have an emergent 
cosmological constant, different from the
one initially chosen to study the dynamics of  black hole interiors. We find that the asymptotic emergence of `charge' and cosmological 
constant that develop after the bounce is purely quantum geometric in origin. The  `charged' Nariai and 
anti-Bertotti-Robinson spacetimes occur in only one side of the temporal evolution in the Kantowski-Sachs spacetime with positive and 
negative cosmological constant and in higher genus black hole interior spacetimes with a negative cosmological constant respectively. The higher genus 
black hole interior with a positive cosmological constant does not yield any of these spacetimes in the asymptotic limit. The emergence of 
`charged' Nariai and anti-Bertotti-Robinson spacetimes present for the first time examples of time asymmetric evolution in LQC, and 
indicate the same for the black hole interiors in the loop quantization. 
However, note that  the uncharged Nariai spacetime which is a non-singular spacetime classically \cite{n1}, can be considered as the 
maximal Schwarzschild-de Sitter black hole where the
cosmological horizon and the black hole horizon of a
Schwarzschild-de Sitter black hole coincide \cite{Bousso}. Thus the emergent spacetimes in the above cases in LQC are closely related to the 
original spacetimes - but are rather special as they are nonsingular and are parameterized by an emergent  `charge' and an emergent  
cosmological constant. Our analysis shows that with a fine tuning of the value of the cosmological constant in the Kantwoski-Sachs 
spacetime, it is possible to obtain `uncharged' Nariai spacetime. However, such a spacetime turns out to be unstable \cite{js3}. Similarly, 
for the higher genus black hole interior, an anti-Bertotti-Robinson spacetime with a vanishing emergent cosmological constant can arise, but 
it too is unstable.

All the above interpretations of emergent spacetime after the bounce is based on the fact that it is a product of two spaces having constant curvature $R^0_0 = R^1_1 = k_1$ and $R^2_2 = R^3_3 = k_2$, which could be written as $k_1=\lambda +\alpha_1, k_2=\lambda+\alpha_2$. Now if we set $\alpha_1=-\alpha_2<0$, it is charged Nariai while for $\alpha_1=-\alpha_2>0$, it is anti-Bertotti-Robinson with a cosmlogical constant. Note that anti-Bertotti-Robinson spacetime has electric energy density negative. The moot point is simply that what emerges after bounce is a product of two constant curvature spaces which by proper splitting of these constants lead to a nice interpretation as a mixture of Nariai and Bertotti-Robinson spacetimes which are exact solutions of classical Einstein equation. It is remarkable that emergent spacetime is solution of classical equation albeit with a proper choice of constants. This may be an innate characteristic of quantum dynamics of 
this type of spacetimes and is perhaps reflection of discreteness in spacetime structure.

The manuscript is organized as follows. In Sec. II, we summarize some of the main features of the classical theory for the Kantowski-Sachs and the higher genus black hole interior 
in terms of the Ashtekar variables and the way the symmetry reduced triads are related to the metric components in these models. In the same section, we introduce the classical Hamiltonian in Ashtekar variables and derive the
classical Hamilton's equations which result in a singular evolution. Note that in the classical evolution of these models, there is no 
emergence of the classical Nariai or anti-Bertotti-Robinson spacetimes. The loop quantum evolution derived from the effective Hamiltonian 
constraints is discussed in Sec. III, which is non-singular. Here after deriving the modified Hamilton's equations, we first discuss the 
boundedness of expansion and shear scalars, and then study the numerical solutions in different cases. In section IV we consider the 
properties of the asymptotic spacetime emerging in the
loop quantum evolution and find them as the `charged' Nariai spacetime and the
anti Bertotti-Robinson spacetime of classical GR. These spacetimes have an emergent `charge' and an emergent cosmological constant, both 
arising from the loop quantum geometry of the spacetime. The values of these quantum geometric
 `charge' and cosmological constant are computed in Sec. V.  We then consider the fine tuned case where there is no `charge' in the spacetime
emerging from the loop quantum model of Kantowski-Sachs spacetime and a vanishing emergent cosmological constant
in the loop quantum model of higher genus black hole interior in Sec. VI. We summarize with a  discussion of the results in Sec. VII.

\section{Kantowski-Sachs and the higher genus black hole interior: classical aspects in Ashtekar variables}
In this section, we summarize the classical Hamiltonian constraint and the resulting Hamilton's equations for the Kantowski-Sachs spacetime \cite{ab,bv,dwc,js1} and the higher genus black hole interior \cite{Brannlund}. We first discuss the case of the Kantowski-Sachs spacetime. Due to the underlying symmetries of this spacetime, which has the spatial topology $\mathbb{R}\times\mathbb{S}^2$,  the Ashtekar-Barbero connection $A^i_a$ and the triad $E^a_i$ simplify to \cite{ab}
\ba
A_a^i \tau_i \d x^a &=& \munderbar{c} \tau_3 \d x + \munderbar{b} \tau_2 \d \theta - \munderbar{b} \tau_1 \sin\theta \d \phi + \tau_3 \cos \theta \d \phi ~,\\
\munderbar{E}_i^a \tau_i \p_a &=& \munderbar{p}_c \tau_3 \sin \theta \p_x + \munderbar{p}_b \tau_2 \sin \theta \p_\theta - \munderbar{p}_b \tau_1 \p_\phi ~.
\ea
Here $\tau_i = - i \sigma_i/2$, and $\sigma_i$ are the Pauli spin matrices. The symplectic structure is determined by
\be
{\bf \Omega}=\frac{L_o}{2G\gamma}\left(2 \d{\munderbar{b}}\wedge \d{\munderbar{p}_b}+d{\munderbar{c}}\wedge \d{\munderbar{p}_c}\right),
\ee
where $L_0$ is the length of fiducial cell along the noncompact radial direction. It is convenient to work with rescaled triads and connections,
\be
p_b=L_o \munderbar{p}_b, \text{  }p_c=\munderbar{p}_c, \text{  }b=\munderbar{b}, \text{  }c=L_o \munderbar{c} \label{scaling} ~
\ee
which are independent of any change to rescaling of the fiducial cell. These variables satisfy the following Poisson brackets:
\be \label{poisson}
\left\lbrace b,p_b \right\rbrace =G\gamma, \qquad \left\lbrace c,p_c \right\rbrace =2G\gamma,
\ee
where $\gamma \approx 0.2375$ is the Barbero-Immirzi parameter whose value is set from the black hole thermodynamics in LQG.

The triads are related to the metric coefficients of the Kantowski-Sachs metric as:
\begin{eqnarray}\label{relationtometric}
 p_b=L_o \sqrt{g_{xx}g_{\theta \theta}}, \qquad p_c = g_{\theta \theta},
\end{eqnarray}
and to the physical volume of the fiducial cell as $V=4 \pi p_b \sqrt{p_c}$. The connection components are proportional to the time derivatives of the
metric components. These relationships are determined from the dynamical equations obtained from the Hamiltonian, whose gravitational part is given by \cite{ab}
\be \label{classical_ks}
\mathcal{H}_{KS}^{(g)}=\frac{-N}{2G\gamma^2} \left[2bc\sqrt{p_c}+\left(b^2+\gamma^2\right) \frac{p_b}{\sqrt{p_c}} \right] .
\ee

Similarly, in the case of the higher genus black hole interior), the Ashtekar-Barbero connection and the conjugate triads simplify to \cite{Brannlund}
\ba
A_a^i \tau_i \d x^a &=& \munderbar{c} \tau_3 \d x + \munderbar{b} \tau_2 \d \theta - \munderbar{b} \tau_1 \sinh\theta \d \phi + \tau_3 \cosh \theta \d \phi ~,\\
\munderbar{E}_i^a \tau_i \p_a &=& \munderbar{p}_c \tau_3 \sinh \theta \p_x + \munderbar{p}_b \tau_2 \sinh \theta \p_\theta - \munderbar{p}_b \tau_1 \p_\phi ~.
\ea
Using the  same rescaling as in \eqref{scaling}, one finds that the rescaled triads $(p_b,p_c)$ and connection components $(b,c)$ are independent of
$L_o$. The symplectic structure, the poisson brackets and the relations between the triads and metric coefficients remain the same as for in the Kantowski-Sachs
spacetime. The physical volume of the fiducial cell for the higher genus black hole interior, differs from that of the Kantowski-Sachs spacetime
by an overall factor and is given by\footnote{In higher genus black hole interior spacetimes the area of constant $t-x$ surface is given by $\int^{2\pi}_0 \int^{\theta_0}_0 \sinh(\theta) \d\theta \d\phi=2 \pi (\cosh(\theta_0)-1)$.}  $V=2 \pi (\cosh(\theta_0)-1) p_b \sqrt{p_c}$. Further, the metric of the spacetime in this case is similar to the Kantwoski-Sachs spacetime albeit which is spatially open, and is same as of the Bianchi-III LRS spacetime. Thus, as the Schwarzschild case, the higher genus black hole interior can also be studied using a homogeneous anisotropic spacetime metric. Due to this reason, in the following discussion
the higher genus black hole interior spacetime in our analysis will also be referred to as Bianchi-III LRS spacetime.

The gravitational part of the classical Hamiltonian for the interior of higher genus black holes in Ashtekar variables turns out to be \cite{Brannlund}
\be \label{classical_bh}
\mathcal{H}_{HG}^{(g)}=\frac{-N'}{2G\gamma^2} \left[2bc\sqrt{p_c}+\left(b^2-\gamma^2\right) \frac{p_b}{\sqrt{p_c}} \right] 
\ee
where we have absorbed the factor $(\cosh(\theta_0)-1)/2$ in to the lapse $N'$. 

Our goal in this manuscript is to study the dynamics of Kantowski-Sachs and higher genus black hole interior spacetimes in the presence of cosmological constant.
Since the forms of the gravitational Hamiltonians for the Kantowski-Sachs and the Bianchi-III LRS spacetime (higher genus black hole interior) are very similar, we can write them together in the following form by setting
$N$ and $N'$ to unity, and add a term corresponding to the cosmological constant. We obtain the classical Hamiltonian as:
\be
\mathcal{H}_{cl}=\frac{-1}{2G\gamma^2} \left[2bc\sqrt{p_c}+\left(b^2+k \gamma^2\right) \frac{p_b}{\sqrt{p_c}} \right] + 4 \pi p_b \sqrt{p_c} \, \rho_\Lambda
\ee
where $k=1$ for the Kantowski-Sachs spacetime  and $k=-1$ for the higher genus black hole interior, and $\rho_\Lambda = \Lambda/8 \pi G$ with $\Lambda$ allowed to have both signs.

 Using the Hamilton's equations, the classical equations of motion for the symmetry reduced connection and triads turn out to be,
\begin{eqnarray}
\dot{p_b}&=&-G\gamma \frac{\partial \mathcal{H}}{\partial b}=\frac{1}{\gamma}\left(c\sqrt{p_c}+\frac{bp_b}{\sqrt{p_c}}\right) \label{p_b}\\
\dot{p_c}&=&-2G\gamma \frac{\partial \mathcal{H}}{\partial c}=\frac{1}{\gamma}2b\sqrt{p_c} \label{p_c}\\
\dot{b}&=&G\gamma \frac{\partial \mathcal{H}}{\partial p_b}=\frac{-1}{2\gamma \sqrt{p_c}}\left( b^2+k\gamma^2 \right) + 4 \pi G \gamma \sqrt{p_c} \rho_\Lambda  \\
\dot{c}&=&2G\gamma \frac{\partial \mathcal{H}}{\partial p_c}=\frac{-1}{\gamma \sqrt{p_c}}\left(bc-\left( b^2+k\gamma^2 \right)\frac{p_b}{2p_c} \right)  + 4 \pi \gamma G p_b \frac{\rho_\Lambda}{\sqrt{p_c}}  \label{ccl}~
\end{eqnarray}
where the `dot' refers to the derivative with respect to proper time.

In the classical theory, unless the matter violates weak energy condition, evolution determined by the above equations generically leads to a singularity. For matter satisfying weak energy condition, there are
two special but
 highly fine tuned cases which lead to a singularity free spacetime.  These two cases are allowed by demanding that $b=0$ at all times for: (i)  the positive cosmological constant for $k=1$, and (ii) the
negative cosmological constant for $k=-1$. The first case leads to the classical uncharged Nariai spacetime which is topologically $dS_2 \times \mathbb{S}^2$, which is discussed in Ref. \cite{dwc}. It is straightforward
to see from the above equations that the second case leads to  the uncharged anti Nariai spacetime which has a topology $AdS_2 \times \mathbb{H}^2$.
These spacetimes are non-singular in the classical theory \cite{n1}. In  general, when $b=0$ is not assumed, for arbitrary values of
cosmological constant, the evolution is singular. The same is true if instead of $\rho_\Lambda$ one chooses energy density sourced with perfect fluids or matter fields obeying weak energy condition.
In the
backward or the forward evolution from a finite volume at a small spacetime curvature, the behavior of the triads and connections is such that the
spacetime curvature becomes infinite in a finite time when the physical volume approaches zero.

\section{Effective loop quantum dynamics of Kantowski-Sachs and higher genus black hole interior spacetimes}
In the loop quantization, the elementary variables are the holonomies of the connection and the fluxes of the corresponding triads. In the case of
the homogeneous spacetimes, fluxes turn out to be proportional to triads,  thus the key modification, in terms of the variables,
 from the classical to the quantum theory appears
in the usage of holonomies, which are the trigonometric functions of the connection components.  The Hamiltonian constraint expressed in terms of holonomies and then quantized results in a quantum difference equation in the triad representation with discreteness determined by the minimum area in LQG. The physical states obtained as the solutions of the quantum Hamiltonian constraint exhibit non-singular evolution, a result which is a direct manifestation of the underlying quantum geometry. Interestingly, the underlying quantum dynamics can be
captured using an effective Hamiltonian constraint. 
The effective dynamics,
derived from the Hamilton's equations,  has been shown to capture the quantum evolution to an
excellent degree of accuracy in different models at all the scales \cite{aps3,apsv,bp,ap,ps12,num2,rad,num3,b1_madrid,num4}. In the following, we assume the validity of the effective spacetime description for the Kantowski-Sachs and higher genus black hole interior spacetimes.

In the effective spacetime description, the LQC Hamiltonian for Kantowski-Sachs and the higher genus black hole spacetime can be written as \cite{bv,Brannlund}:
\be\label{effham}
\mathcal{H}=\frac{-1}{2G\gamma^2}\left[2\frac{\sin{(b\delta_b)}}{\delta_b}\frac{\sin{(c\delta_c)}}{\delta_c}\sqrt{p_c}+\left(\frac{\sin^2{(b\delta_b)}}{\delta_b^2}+k\gamma^2\right)\frac{p_b}{\sqrt{p_c}}\right] \, + \, 4 \, \pi  p_b \sqrt{p_c} \rho_\Lambda,
\ee
where $k = + 1$ and $-1$ for the Kantowski-Sachs spacetime and the Bianchi-III LRS/higher genus black hole spacetime. Here $\delta_b$ and $\delta_c$ are the functions of triads, whose exact form is dictated by the
loop quantization. For the quantization corresponding to the improved dynamics prescription in LQC \cite{aps3}, these are given by
\be
\delta_b=\sqrt{{\Delta}} \, \frac{1}{{p_c^{1/2}}}, \quad \delta_c={\sqrt{\Delta}} \, \frac{p_c^{1/2}}{p_b} \label{imdy}
\ee
where $\Delta = 4 \sqrt{3} \pi \gamma \lp^2$ corresponds to the minimum area eigenvalue in LQG. Recently, this quantization prescription
has been shown to be the unique choice which leads to physical predictions independent of the choice of the fiducial length $L_o$, and yield universal bounds on expansion and shear scalars
for the geodesics in the effective spacetime description in the Kantowski-Sachs model for arbitrary matter \cite{js1}. These properties, which are not shared by other possible quantization prescriptions,
also hold true for the Bianchi-III LRS spacetimes as shown later in this section.

To obtain the effective dynamics, we first obtain the Hamilton's equations from (\ref{effham}) which are then numerically solved. The resulting Hamilton's equations are:
\ba
\dot{p_b}&=&\frac{p_b \cos{(b\delta_b)}}{\gamma \sqrt{\Delta}} \left( \sin{(c\delta_c)} + \sin{(b\delta_b)} \right), \label{pb-eff}\\
\dot{p_c}&=&\frac{2p_c}{\gamma \sqrt{\Delta}} \sin{(b\delta_b)} \cos{(c\delta_c)} \label{pc-eff} \\
\dot{b}&=&\frac{c p_c}{p_b \gamma \sqrt{\Delta}}\sin{(b\delta_b)}\cos{(c\delta_c)} \label{b-eff}\\
\dot{c}&=&\frac{1}{\gamma \sqrt{\Delta}}\left(\cos{(b \delta_b)}\left(\sin{(b \delta_b)}+\sin{(c \delta_c)}\right)\frac{b p_b}{p_c}-c \sin{(b \delta_b)} \cos{(c \delta_c)}\right)+\frac{\gamma k p_b}{p_c^{3/2}} ~. \label{c_eff}
\ea
where the `dot' refers to the derivative with respect to the proper time $\tau$, and we have used the Hamiltonian constraint ${\cal H}\approx 0$ to simplify the equations.

In a similar way, modified Hamilton's equations can be derived for arbitrary matter energy density. Unlike the classical theory, the solutions from these equations are non-singular. Various properties of
singularity resolution for the vacuum case were studied in Refs. \cite{ab,bv}. The case of the
massless scalar field for $k=1$ is studied in Ref. \cite{dwc}, and the case of $k=-1$ was earlier investigated in Ref. \cite{Brannlund}. General properties of Kantowski-Sachs model and the issues of singularities are discussed in \cite{js1}, and details of the evolution for different types of matter are studied in Ref. \cite{js3}. In particular, there are no divergences in expansion and shear scalars in the Kantowski-Sachs model, thus pointing towards the nonsingular nature of the spacetime \cite{js1}.
This result can be easily generalized to the higher genus black hole interior as well, so it is worthwhile to discuss it further.

The expansion of a geodesic congruence can be written in terms of triads as
\begin{equation}
 \theta=\frac{\dot{V}}{V}=\frac{\dot{p}_b}{p_b}+\frac{\dot{p}_c}{2p_c}.
\end{equation}
where the derivative is with respect to proper time. Thus for the expansion scalar to be bounded, $\frac{\dot{p}_b}{p_b}$ and $\frac{\dot{p}_c}{p_c}$ have to be bounded. From eqs.(\ref{pb-eff}) and (\ref{pc-eff}), we note
 that the relative rate of change of triads with respect to proper time is the same
for Kantowski-Sachs spacetime and the higher genus black hole interior (the difference due to the sign of curvature affects only the $\dot{c}$ equation). Using these equations, the
expansion scalar can  be obtained as
\begin{equation}
 \theta=\frac{1}{\gamma \sqrt{\Delta}}\left(\sin{(b\delta_b)} \cos{(c\delta_c)} +
         \sin{(c\delta_c)} \cos{(b\delta_b)} + \sin{(b\delta_b)} \cos{(b\delta_b)}\right)
\end{equation}
The dependence of expansion scalar on phase space variables is only through bounded functions.
 Thus there is a maxima that expansion scalar can reach, corresponding to saturation of the
trigonometric terms. Thus the expansion scalar in the Kantowski-Sachs or the higher genus black hole interior has a universal bound
given as
\begin{equation}
 |\theta|\leq\frac{3}{2\gamma\sqrt{\Delta}}\approx \frac{2.78}{l_{pl}},
\end{equation}
where we have used $\gamma \approx 0.2375$.
The shear scalar which signifies the anisotropy seen by an observer following a geodesic congruence in Kantowski-Sachs or the higher genus black hole interior spacetimes can be written as
\begin{equation}
 \sigma^2=\frac{1}{2}\sum_{i=1}^{3}\left(H_i-\frac{1}{3}\theta \right)^2=\frac{1}{3}\left(\frac{\dot{p}_c}{p_b}-\frac{\dot{p}_b}{p_b}\right)^2,
\end{equation}
where $H_i = \dot a_i/a_i$ are the directional Hubble rates. Using eqs.(\ref{pb-eff}) and (\ref{pc-eff}), we obtain
\begin{equation}
 \sigma^2=\frac{1}{3\gamma^2\Delta}\left(2\sin{(b\delta_b)}\cos{(c\delta_c)}-\cos{(b\delta_b)} \left(\sin{(c\delta_c)}+\sin{(b\delta_b)} \right)\right)^2 ~.
\end{equation}
As the expansion scalar, the shear scalar is also   bounded with a universal bound \cite{js1}
\begin{equation}
 |\sigma|^2\leq\frac{5.76}{l_{pl}^2}.
\end{equation}
The boundedness of both expansion and shear scalars point towards the geodesic completeness of the
loop quantum model of Kantowski-Sachs and the higher genus black hole interior in the improved dynamics prescription (\ref{imdy}). These 
results are in synergy with similar results is isotropic and Bianchi models in LQC \cite{ps09,cs3,ps-fv,ps11,gs1,ps-we}.

\subsection{Kantowski-Sachs spacetime with a positive cosmological constant}
\begin{figure}[tbh!]\label{ksfigpositive}
\hskip-0.3cm
\includegraphics[width=0.42\textwidth]{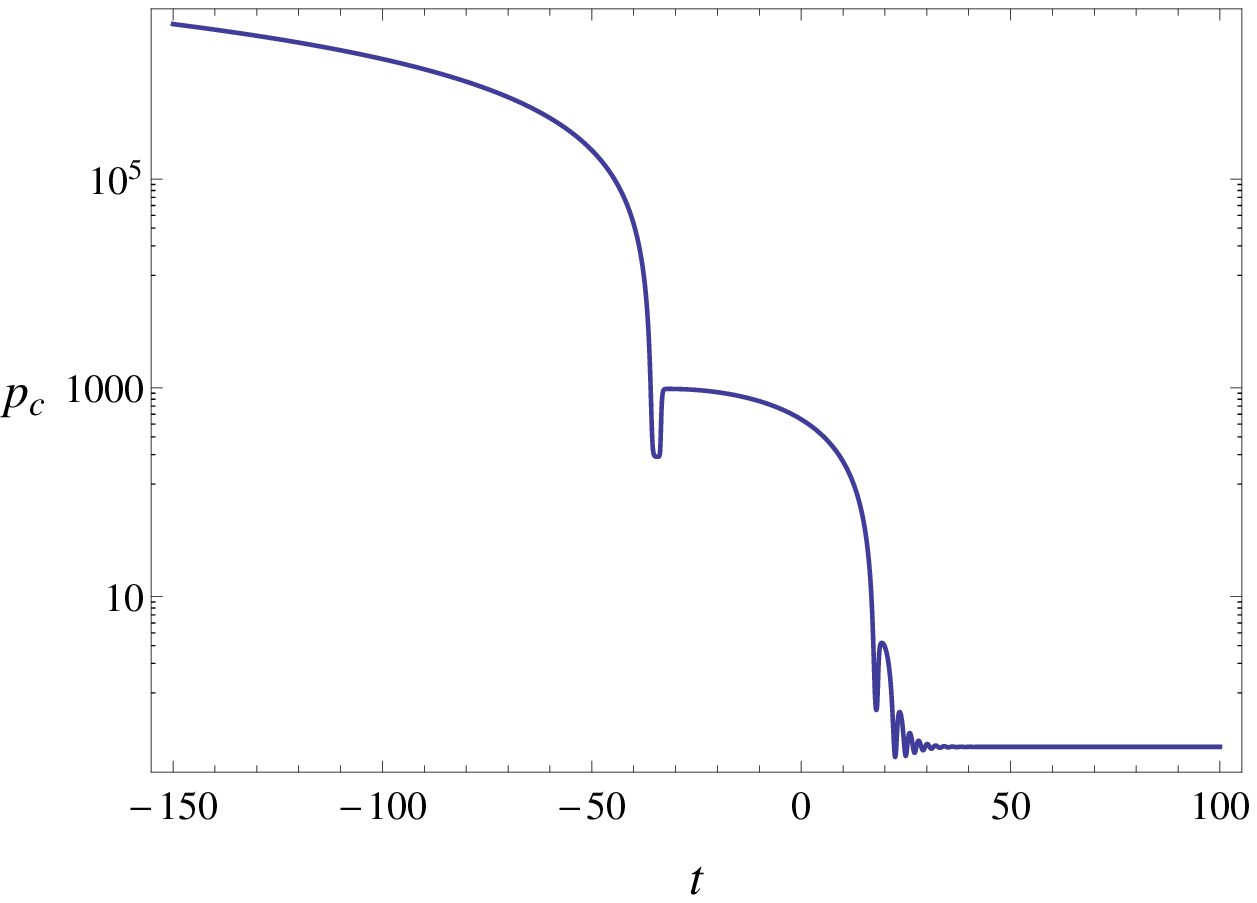}
\hskip0.3cm
    \includegraphics[width=0.42\textwidth]{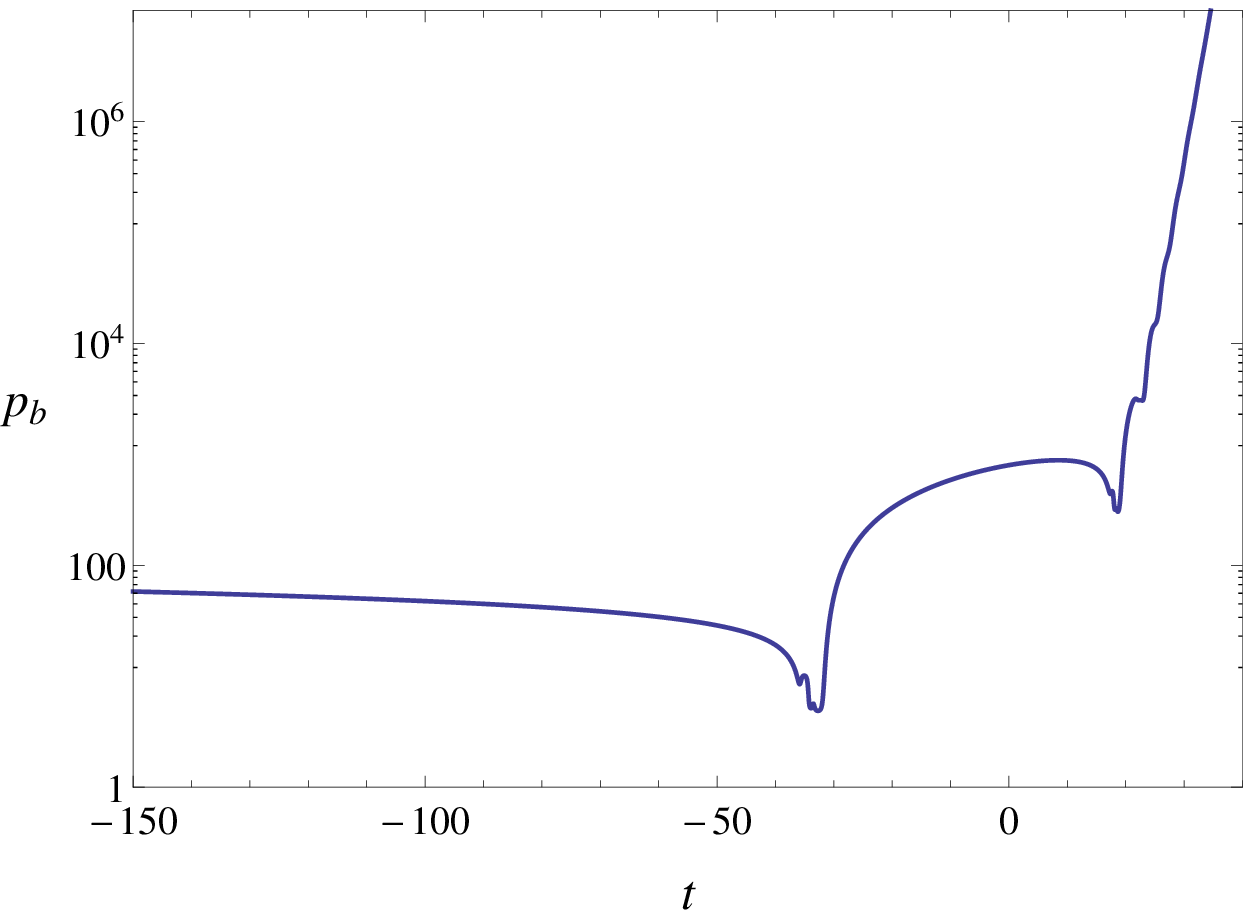}

\vskip0.3cm
\includegraphics[width=0.4\textwidth]{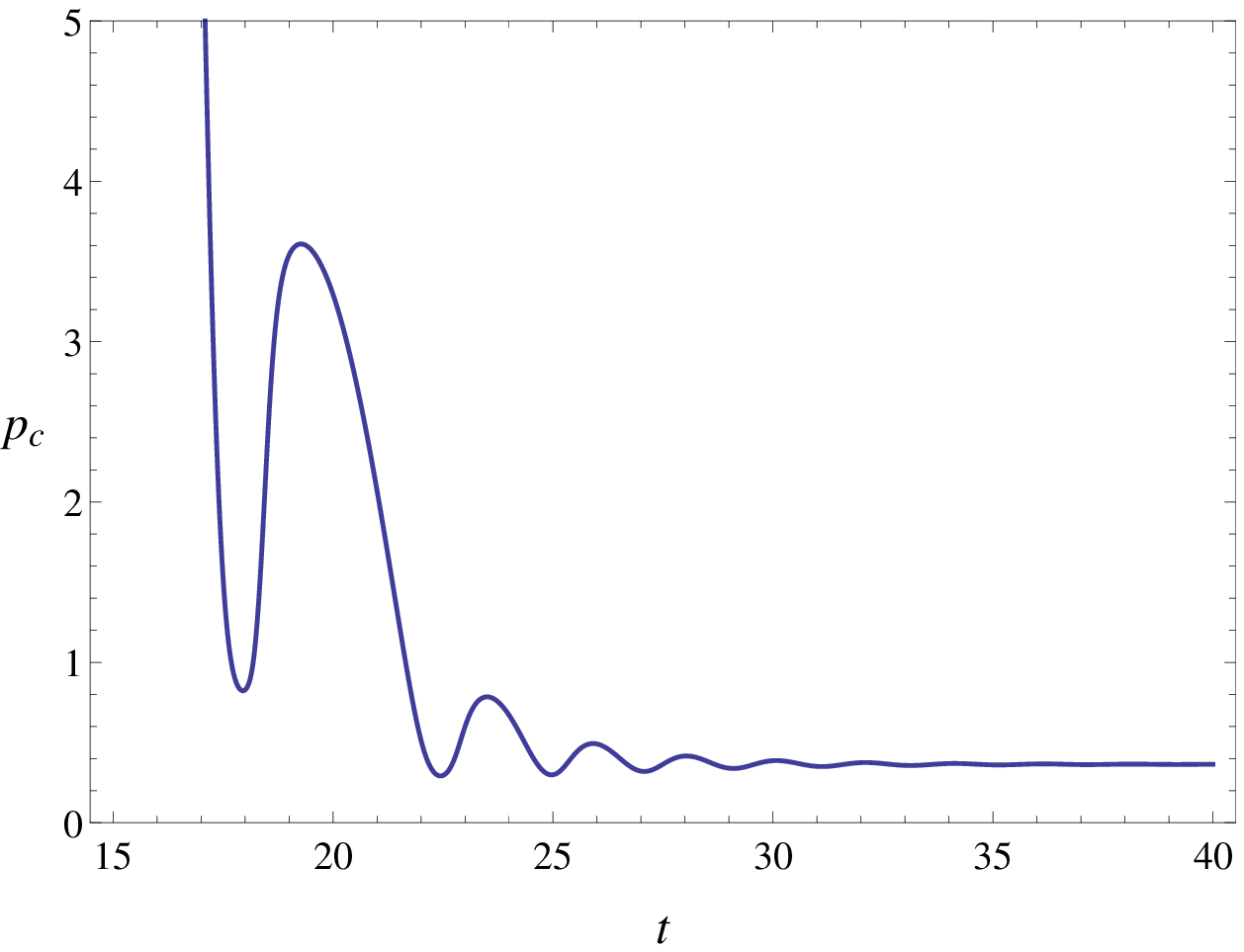}
\hskip0.3cm
\includegraphics[width=0.42\textwidth]{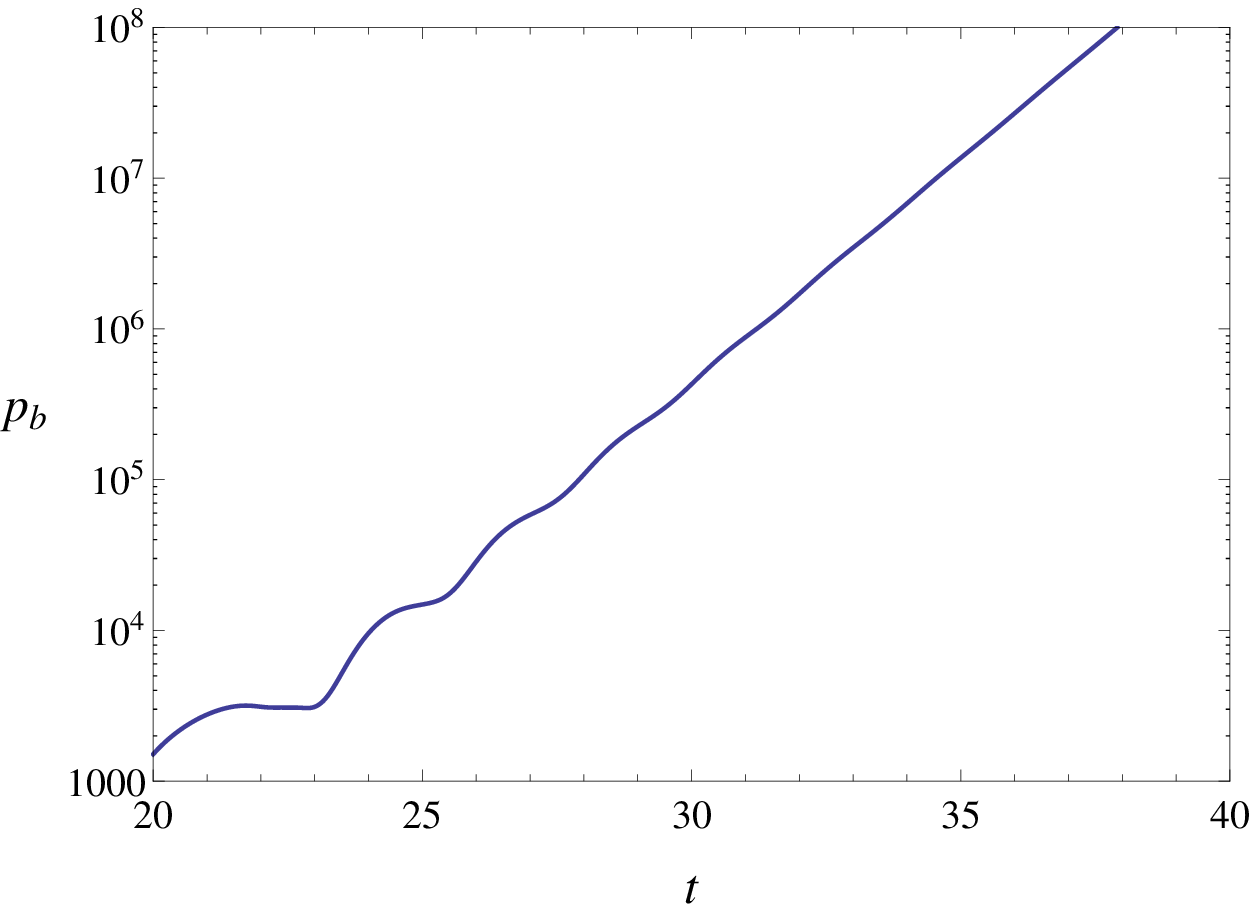}
%

\caption{Evolution of triads vs time for Kantowski-Sachs spacetime with positive cosmological constant is shown.
The initial conditions are chosen at $t=0$ as $p_b(0)=8 \times 10^2$, $p_c(0)=5 \times 10^2$, $b(0)=-0.15$,
$\rho_{\Lambda}(0)=10^{-7}$. (All values in the numerical simulations are in Planck units). Initial value of connection component $c$ is obtained from the vanishing of the Hamiltonian constraint.
The top plots show the behavior in the forward and the backward evolution, where as the bottom plots show the zoomed in behavior of
$p_b$ and $p_c$ in the future evolution for positive time. We find that $p_c$ approaches a constant value asymptotically. }

\end{figure}
\begin{figure}[!ht]
  \centering
    \includegraphics[width=0.48\textwidth]{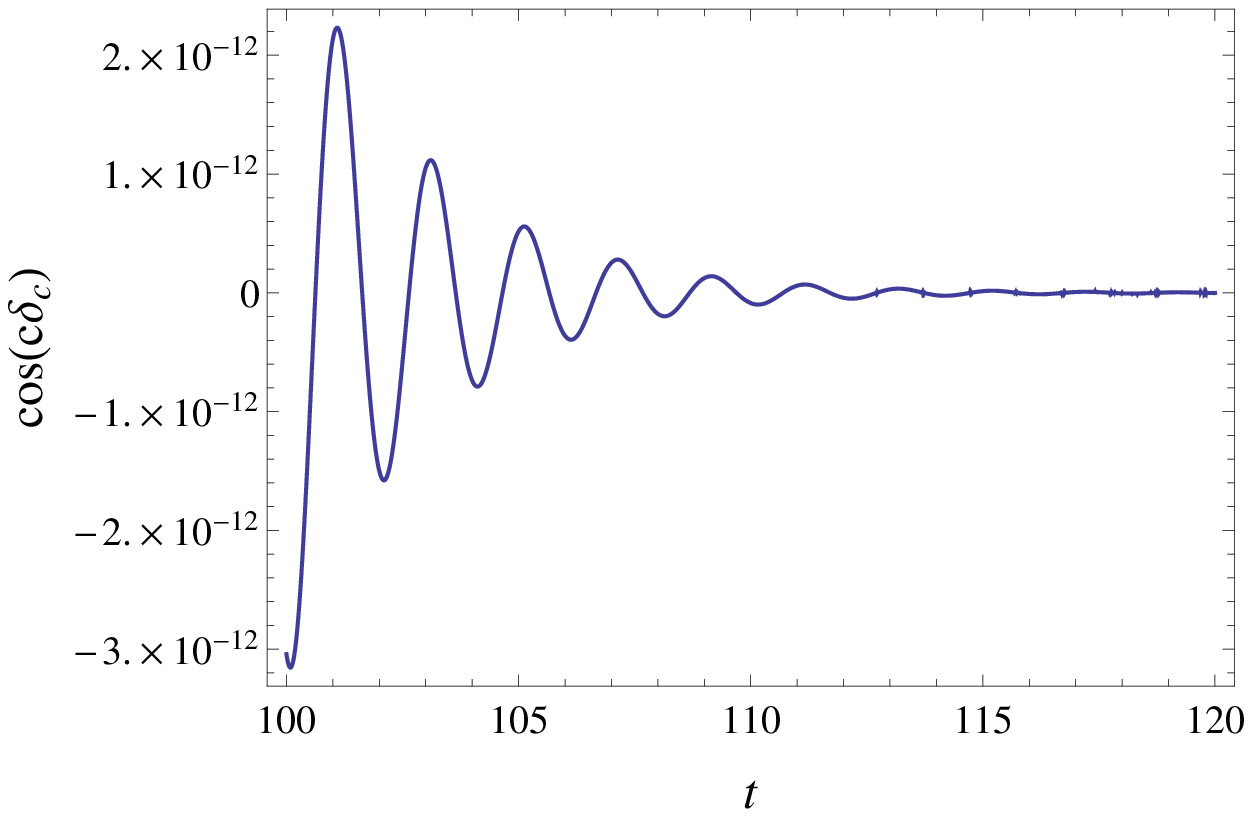}
    \includegraphics[width=0.45\textwidth]{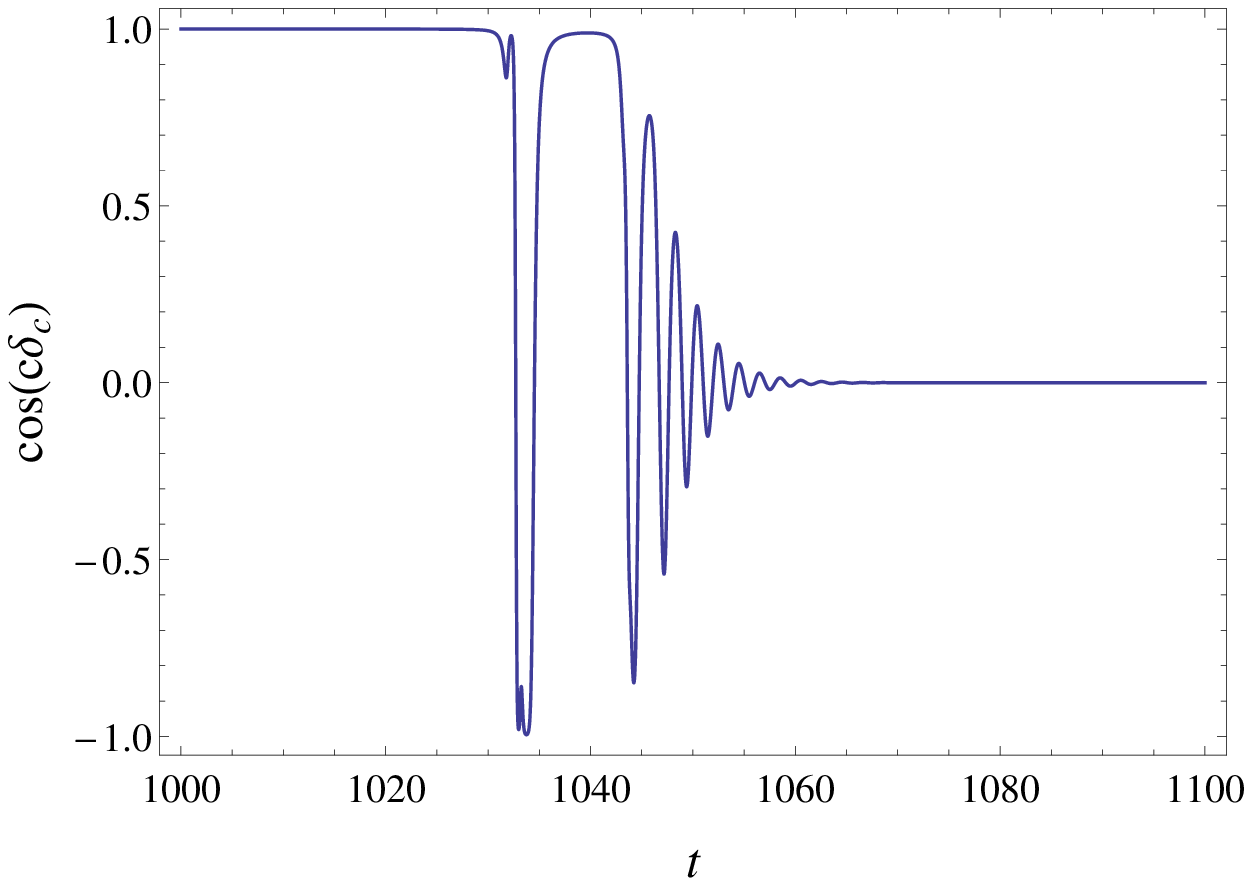}
\caption{The behavior of $\cos(c \delta_c)$ is shown in the asymptotic regime in Kantowski-Sachs spacetime where $p_c$ is a constant. The left plot shows the case for the positive
cosmological constant, and the right plot shows for the negative cosmological constant. The initial conditions correspond respectively to those in Fig. 1 and Fig. 3. }
\label{cos1fig}
\end{figure}

We now discuss the results from the numerical simulations performed for positive cosmological constant using the effective dynamics
of Kantowski-Sachs model. As discussed earlier, in this case the classical dynamics in general leads to a singularity where $p_b$ and  $p_c$ vanish,
and the spacetime curvature diverges. In LQC, the evolution is strikingly different. Starting from a large value of triad components and a small spacetime curvature, we find that
$p_b$ and $p_c$ undergo bounces due to quantum gravitational modifications in the effective dynamics.
The triads have different asymptotic behaviors as $t \rightarrow \pm \inf$. In all the numerical simulations that we carried out with a value of cosmological constant not greater than 0.1 in Planck units,
it was observed that $p_c$ expands exponentially in the asymptotic regime, after few bounces,
in one of the directions in time whereas it reaches a constant value in the other \footnote{For close to Planckian values of cosmological constant this may not be true.
For rather large values of $\Lambda$, say greater than 0.1 (in Planck units), one finds de Sitter expansion after quantum bounce in both the directions of time.}.
This behavior is shown in the top left plot of Fig. 1. In the
forward as well as the backward evolution, the triad $p_b$  undergoes exponential expansion in proper time in the asymptotic regimes. The rate of exponential expansion
for $p_b$ in both of the asymptotic regimes turns out to be different. Interestingly, in the asymptotic
region in which $p_c$ is exponentially expanding, $p_b$ has the same rate of expansion as $p_c$ and this rate approaches a constant value.
In this regime, the holonomy corrections are negligible and the
spacetime behaves as a classical spacetime with a small spacetime curvature which is a solution of the classical
Hamilton's equations for the Kantowski-Sachs model, eqs.(\ref{p_b}-\ref{ccl}) with a positive cosmological constant.
 The exponential behavior of the triads
shows that the classical spacetime in this regime is a de Sitter spacetime.
The other side of the temporal evolution, i.e. when for positive values of time, leads to a
spacetime not having small spacetime curvature even long after the bounce. In this region, shown in the large $t$ range of bottom plots in Fig. 1,
while $p_c$ takes a constant value in the asymptotic regime, $p_b$ grows exponentially. The holonomy corrections are large,
denoting that the quantum effects are very significant in this regime. This is evident from the left plot in Fig. \ref{cos1fig}, which shows that
$\cos(c \delta_c)$ approaches zero, and hence $|\sin(c \delta_c)|$ is unity. In the same regime, $b$ is also finite and non-vanishing, and takes a constant value.
It turns out that this asymptotic regime is {\it{not}} a solution of the classical
Hamilton's equations of the Kantowski-Sachs model (eqs.(\ref{p_b}-\ref{ccl})), which do not allow $p_c$ to be a constant when $b$ is non-vanishing.
A more detailed characterization of this region,  will be carried out in the next section. As we will show,
 the loop quantum spacetime in this regime is a product of constant curvature spaces  with an effective metric which is a solution of Einstein field equations for a `charged' Nariai spacetime.

\subsection{Kantowski-Sachs spacetime with a negative cosmological constant}
\begin{figure}[!ht]
    \includegraphics[width=0.47\textwidth]{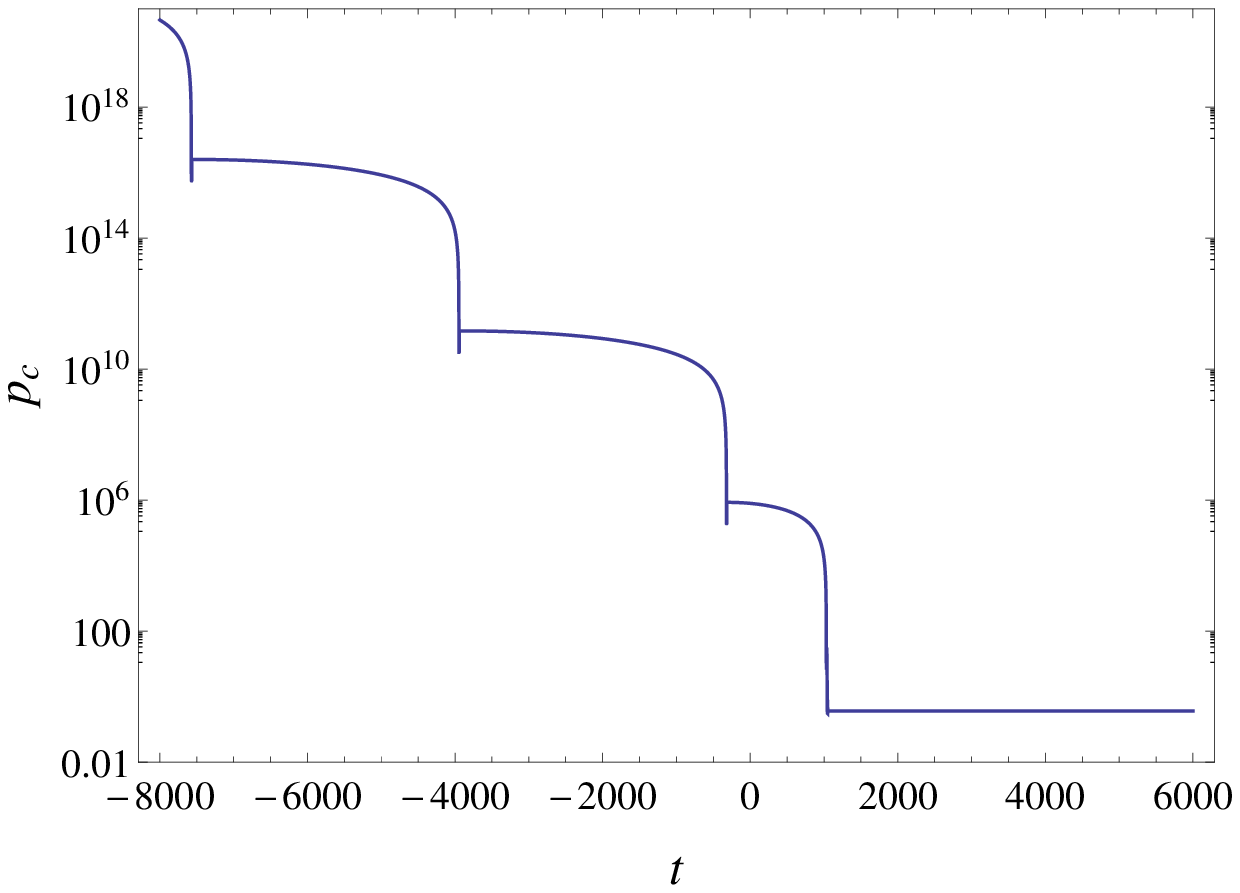}
    \includegraphics[width=0.47\textwidth]{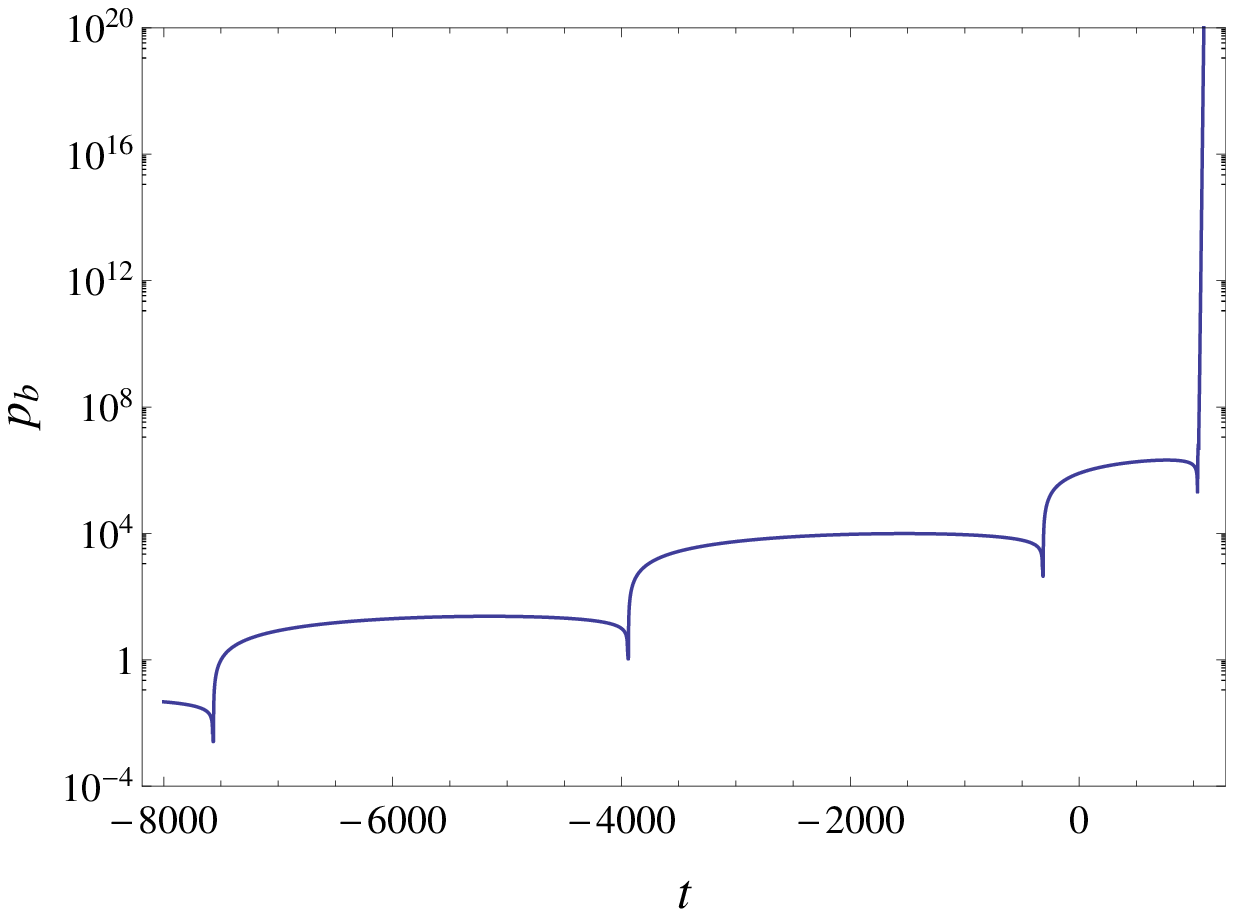}

    \includegraphics[width=0.47\textwidth]{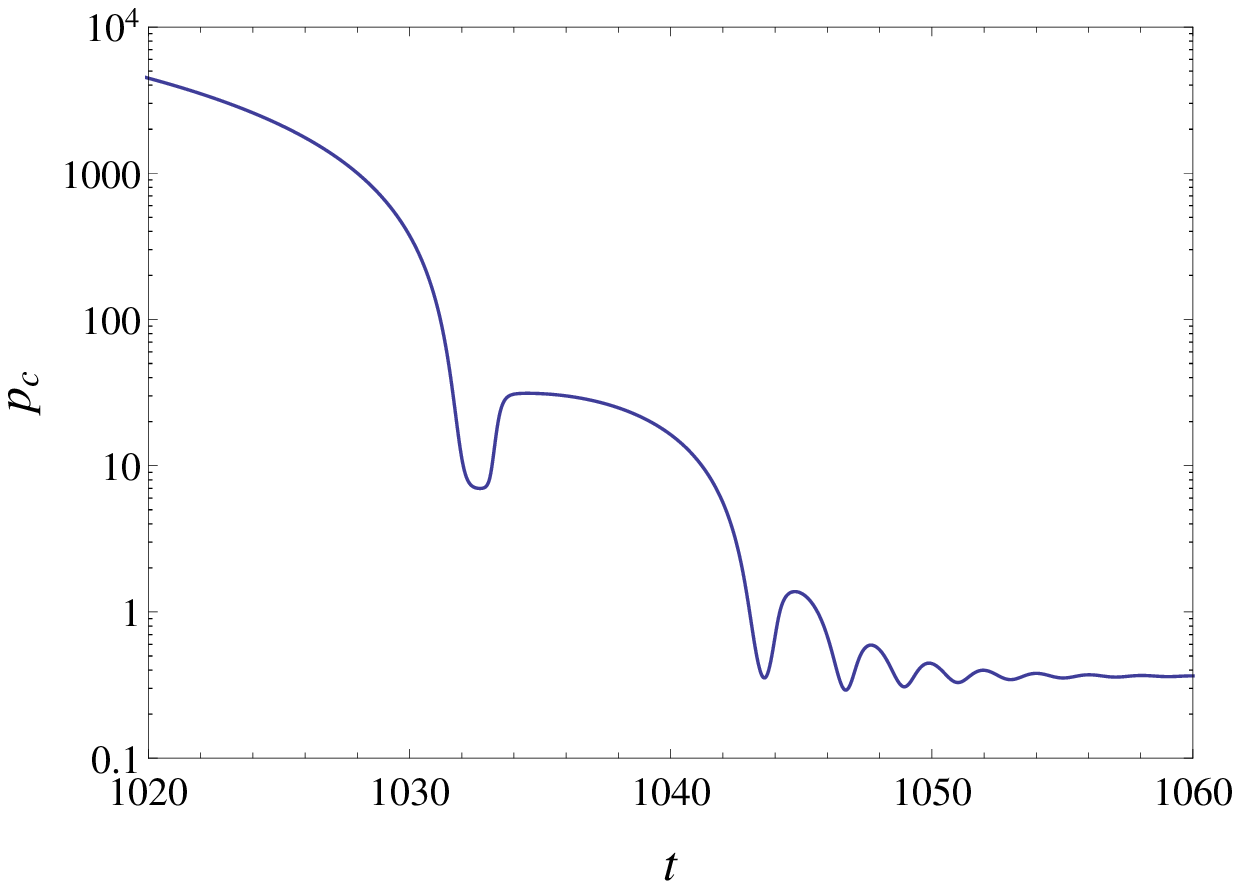}
    \includegraphics[width=0.47\textwidth]{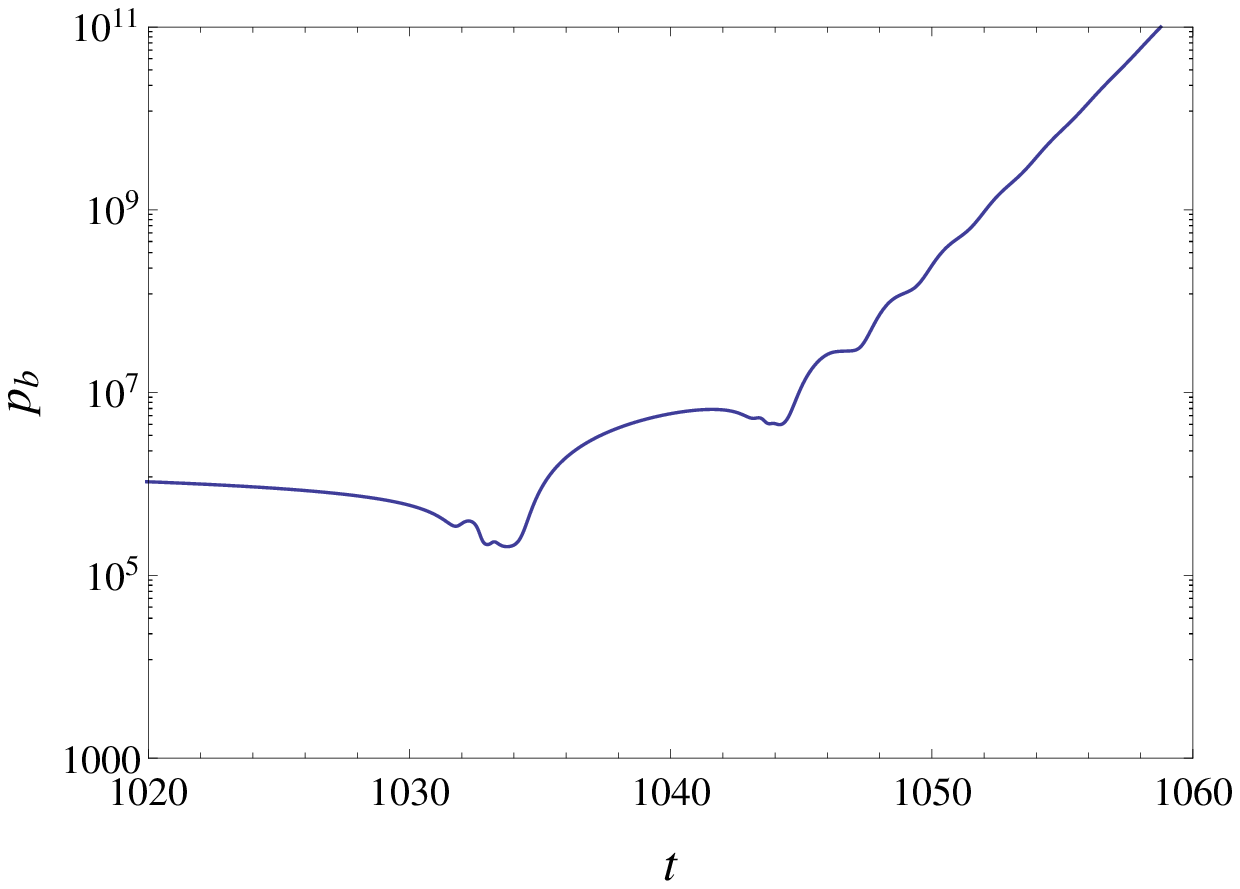}
\caption{These plots show the behavior of triads in time  for the effective dynamics in Kantowski-Sachs sourced with a negative cosmological constant.
The initial conditions are chosen at $t=0$ as $p_b(0)=8 \times 10^5$, $p_c(0)=8 \times 10^5$, $b(0)=-0.05$,
$\rho_{\Lambda}(0)=-10^{-8}$ (in Planck units). For these initial conditions, $p_c$ approaches an asymptotic value in the future time evolution while $p_b$ increases exponentially. The lower plots show the zoomed behavior of upper plots. Existence of such a asymptotic regime, in
either backward or forward evolution is generic for arbitrary initial conditions. }
\label{triads}
\end{figure}

In the presence of the negative cosmological constant, the mean volume in the Kantowski-Sachs spacetime undergoes a recollapse at large scales in the classical theory when
the expansion is halted by the negative energy density pertaining to the cosmological constant. Due to the recollapse, the classical Kantowski-Sachs spacetime
with a negative cosmological constant in general encounters a past as well as
a future singularity with triads vanishing in a finite time. As a result of the quantum gravitational effects
originating from the holonomy modifications in the effective Hamiltonian constraint,
the classical singularity is avoided in the loop quantum dynamics,
generically, as is in the case for the positive cosmological constant. In contrast to the
latter case, the effective dynamics in LQC involves several cycles of bounces and recollapses. Thus, the evolution is cyclic. The behavior of the triads in each cycle
is such that  one triad grows whereas the other decreases. The physical volume
oscillates between fixed maxima and minima thanks to the fixed potential due to the negative cosmological constant. An example of a typical evolution is shown in Fig. \ref{triads}, where
for the considered initial conditions, $p_c$ increases through multiple bounces in the backward evolution whereas $p_b$
decreases. The behavior of the triads in this regime is such that the mean volume undergoes periodic cycles of expansion and contraction.
 In contrast, for the evolution in positive time, the series of bounce and recollapses damps down and $p_c$ approaches a constant value in the
asymptotic regime. In the same regime, $p_b$ grows exponentially. As in the case of the positive cosmological constant, holonomy corrections
are significant in this regime, as can be seen in the right plot in
Fig. \ref{cos1fig}. The regime where $p_c$ approaches a constant value asymptotically is not a solution of the classical Hamilton's equations for the Kantowski-Sachs
spacetime. Further details of this asymptotic solution are discussed in Sec. IV.

\subsection{Bianchi-III LRS spacetime with a negative cosmological constant}

\begin{figure}[!ht]
    \includegraphics[width=0.45\textwidth]{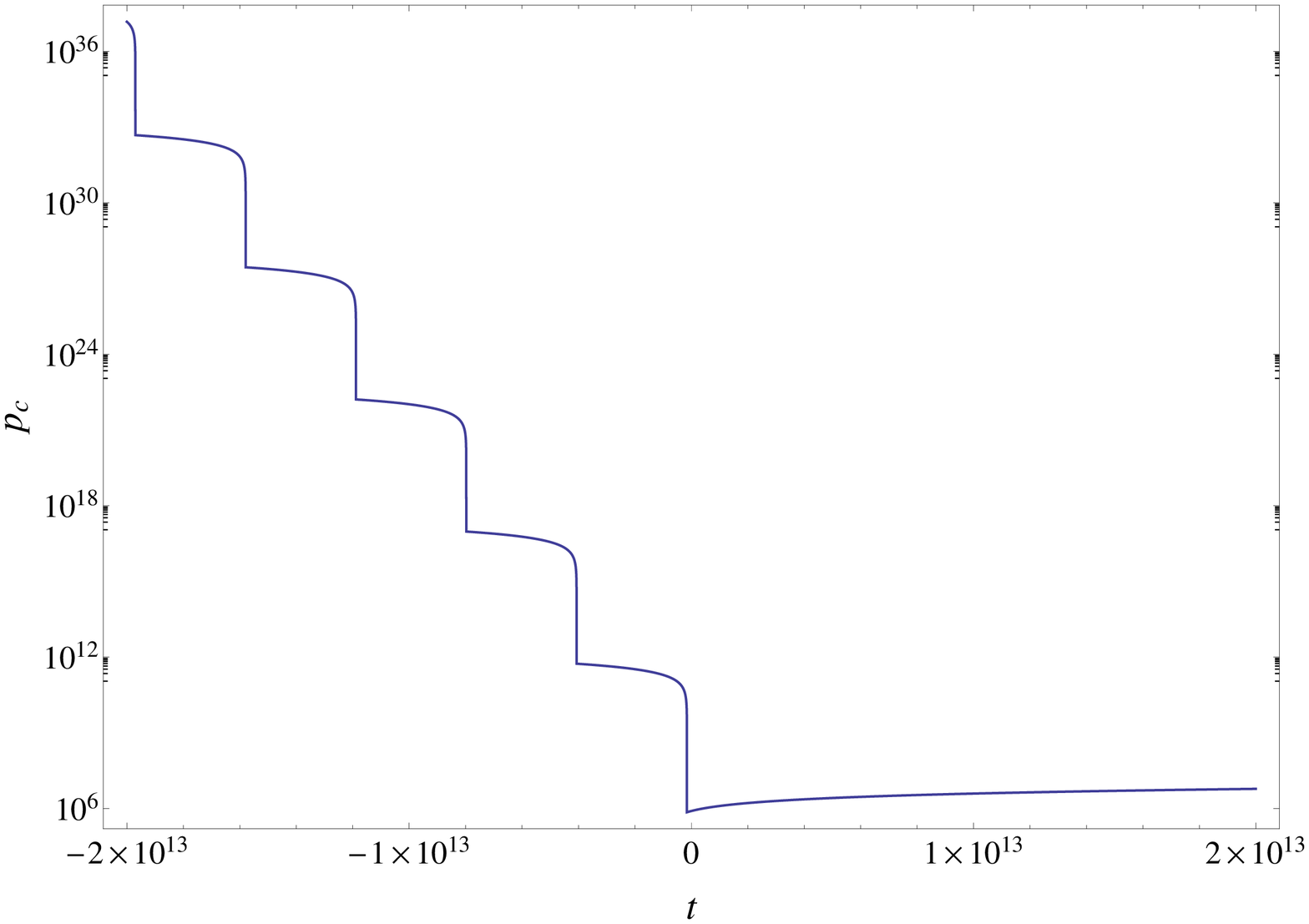}
    \includegraphics[width=0.5\textwidth]{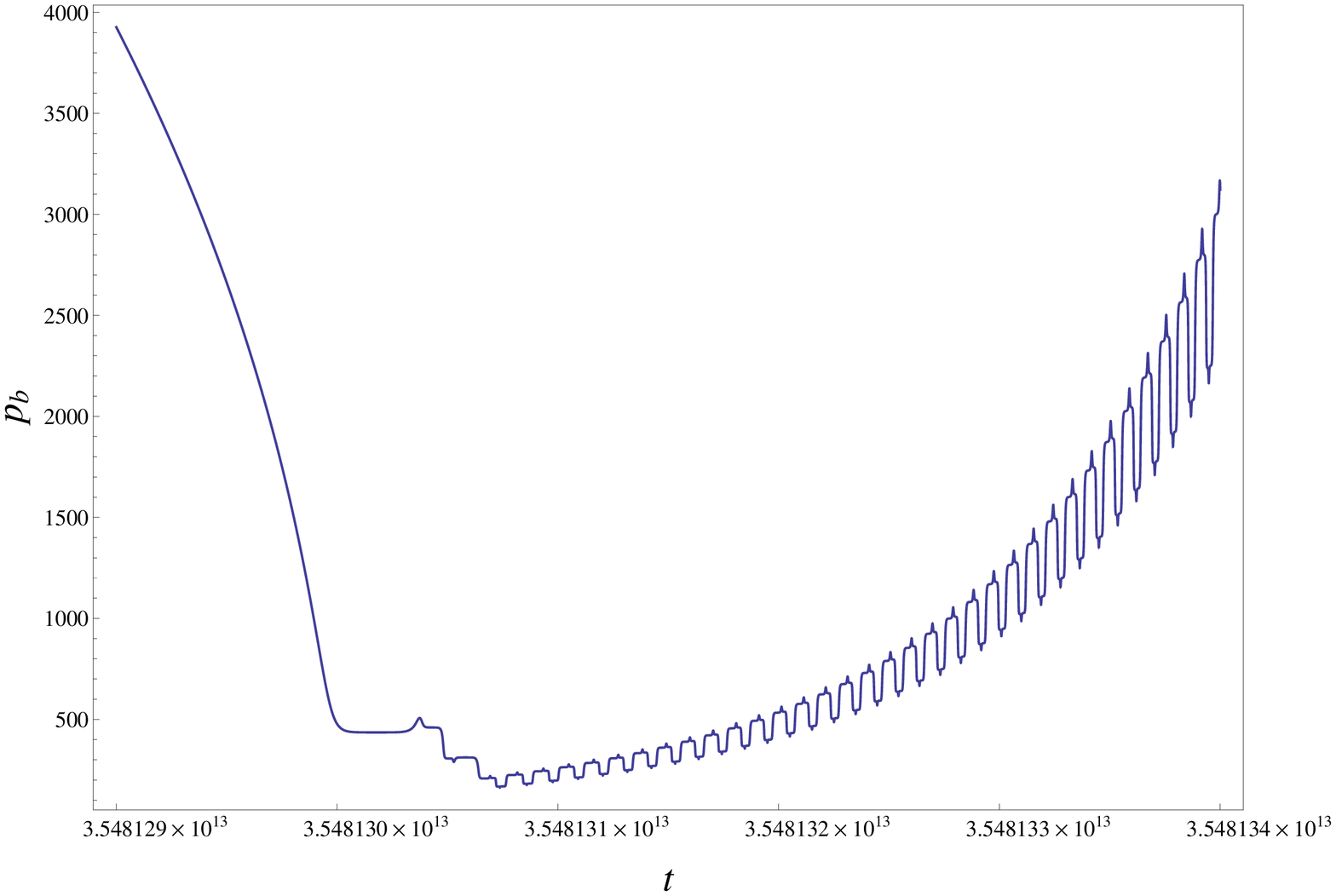}
    \includegraphics[width=0.45\textwidth]{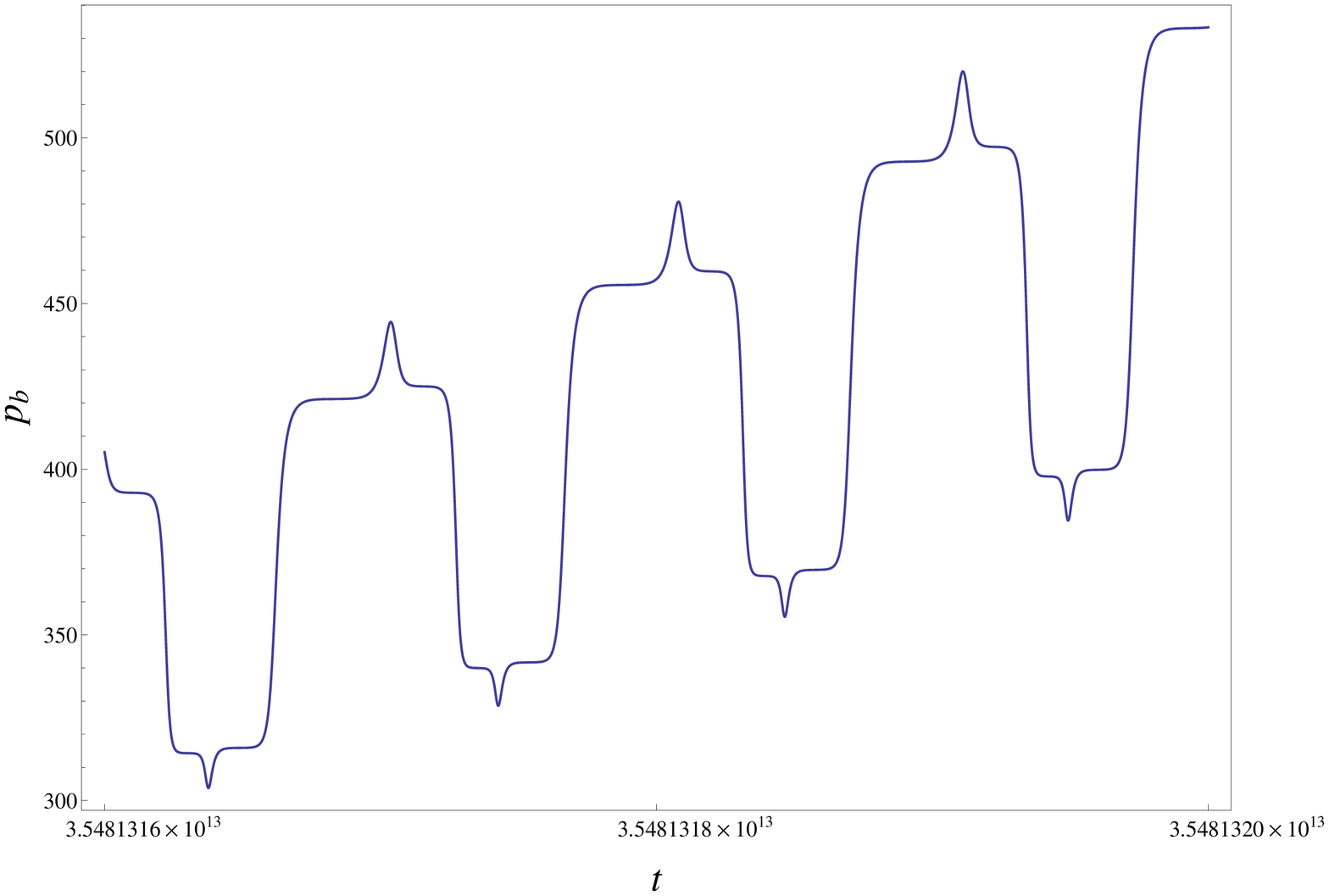}
    \includegraphics[width=0.5\textwidth]{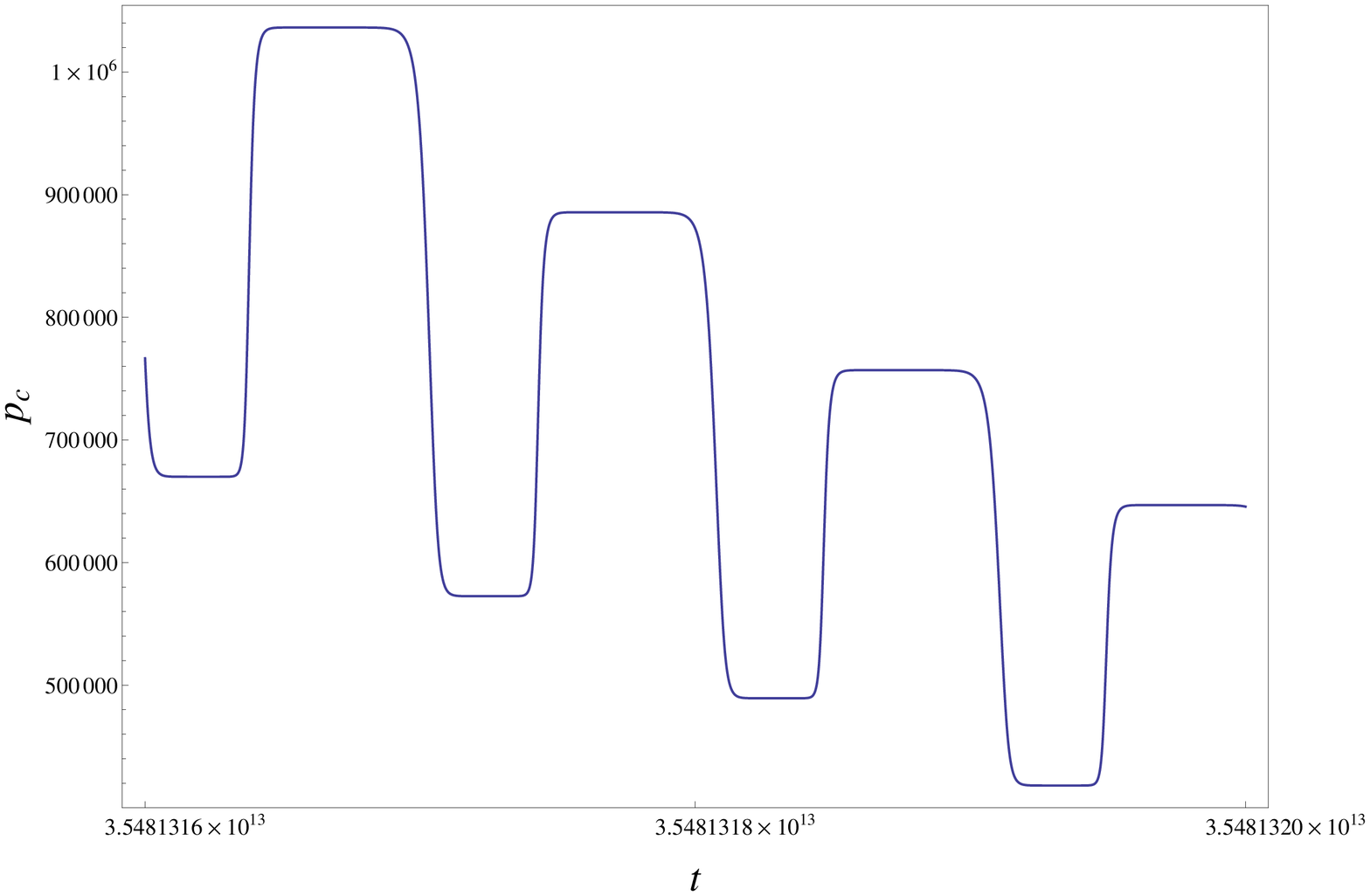}
\caption{These plots show the evolution of triads with respect to the coordinate time for negative cosmological constant in
higher genus black hole spacetime. The lapse for the simulations of this case is chosen to be $N=1/p_b \sqrt{p_c}$. 
The initial conditions are chosen at $t=0$ as $p_b(0)=8 \times 10^5$, $p_c(0)=8 \times 10^5$, $b(0)=0.05$,
$\rho_{\Lambda}(0)=-10^{-8}$ (in Planck units). Evolution is performed  with lapse  $N = 1/p_b \sqrt{p_c}$. The lower plots show some of the 
details in the upper plots.}
\label{triads_bh}
\end{figure}

\begin{figure}[!ht]
  \centering
    \includegraphics[width=0.47\textwidth]{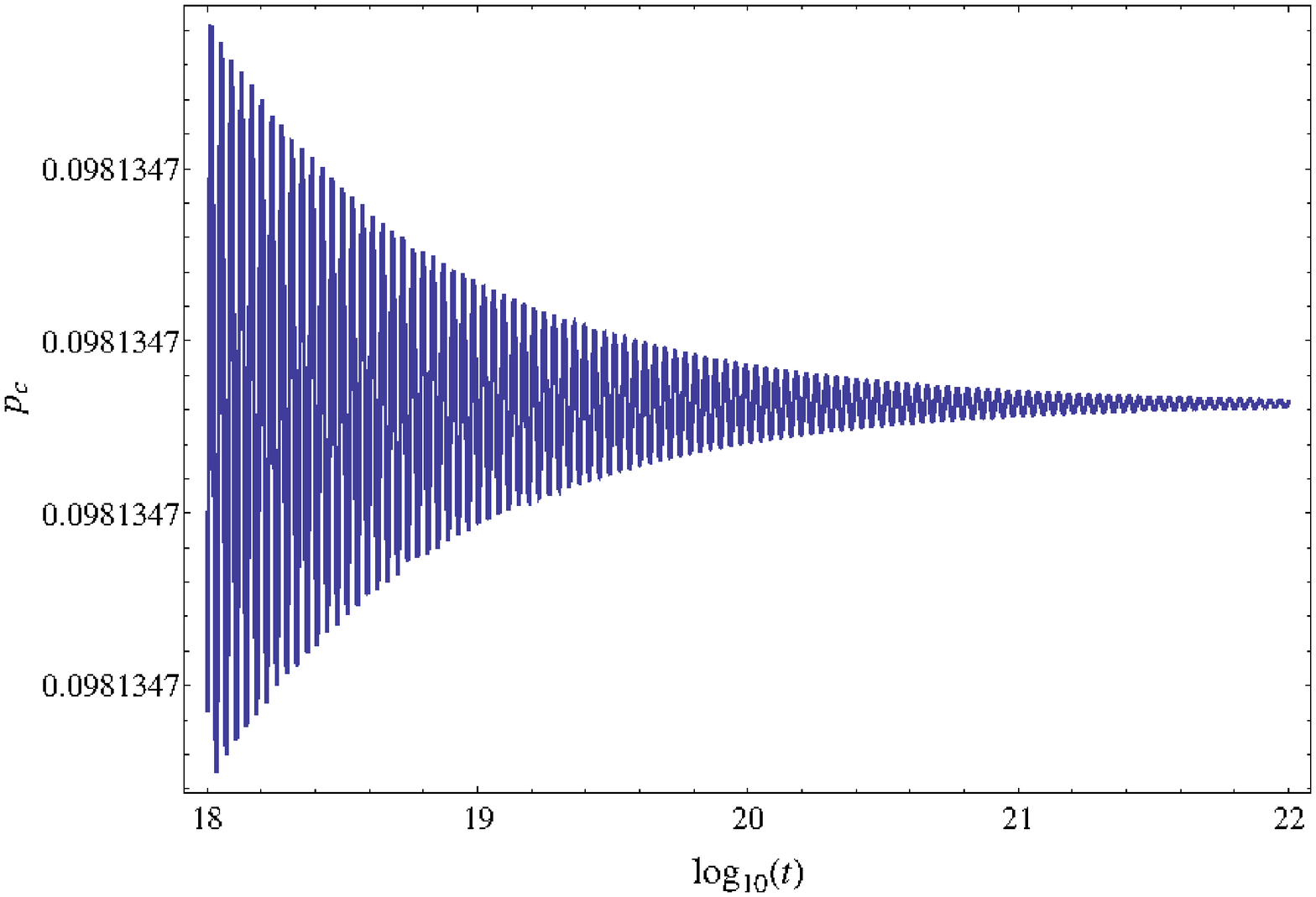}
    \includegraphics[width=0.47\textwidth]{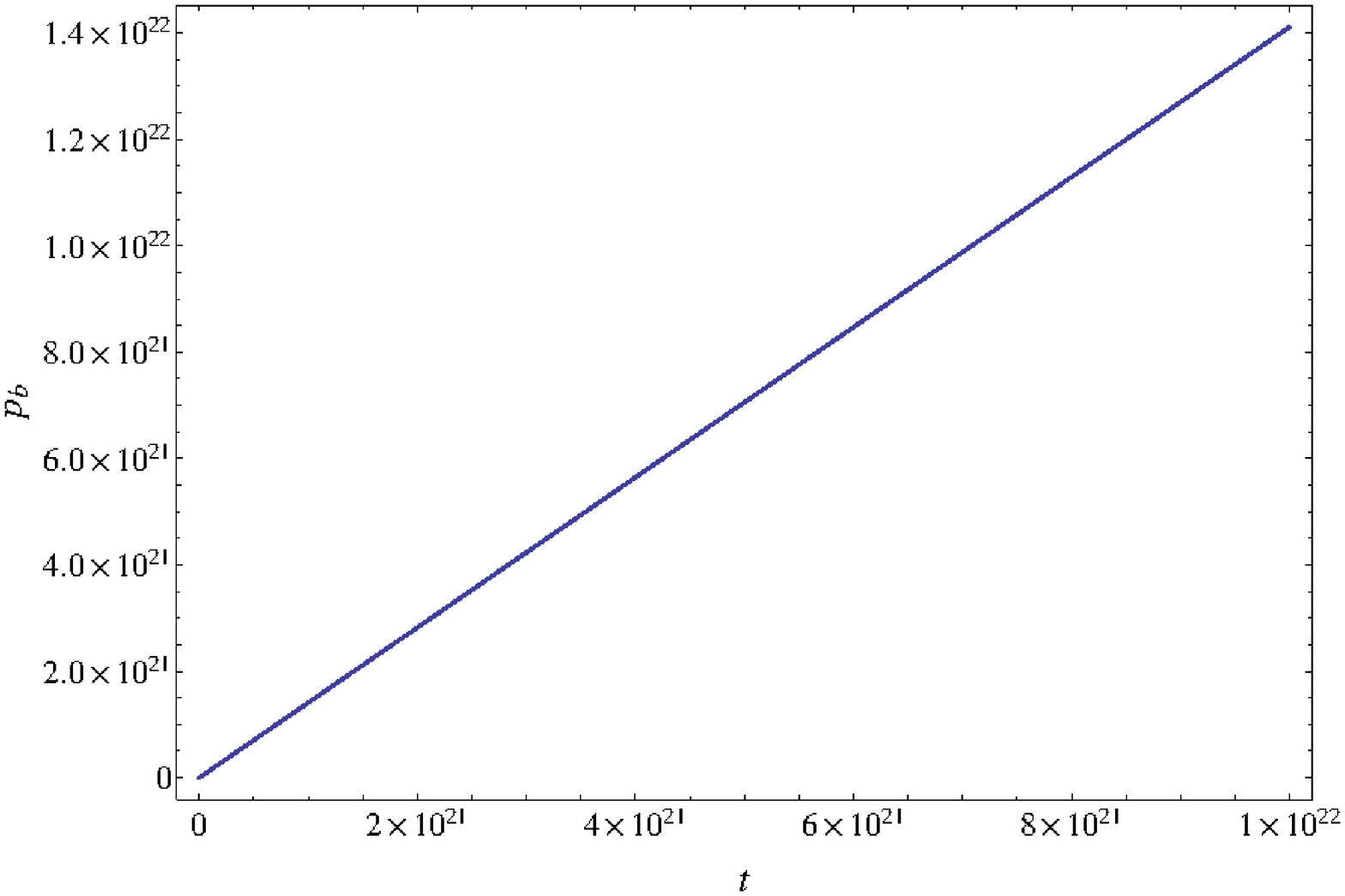}
\caption{The behavior of triads for the higher genus black hole interior is shown in the asymptotic regime.
Triad $p_c$ is constant whereas $p_b$ has a linear growth (corresponding to an exponential
behavior with respect to proper time).  The initial conditions are same as in Figure \ref{triads_bh}.}
\label{_bh_asym}
\end{figure}

The evolution of triads in the Bianchi-III LRS spacetime/higher genus black hole interior with a negative cosmological constant is cyclic in nature, similar
to the case earlier discussed for the Kantowski-Sachs spacetime in Sec. IIIB.
An earlier study of this spacetime in LQC was performed in
Ref. \cite{Brannlund}, where its non-singular properties were first noted. It turns out that to carry out the numerical simulations in this 
case, it is more convenient to work with a lapse $N=1/p_b \sqrt{p_c}$. 

For any given choice of
initial conditions, on one side of the temporal evolution,
the triad $p_c$ undergoes
several bounces and recollapses until it approaches an asymptotic value. In this regime,
 $p_b$ increases linearly in coordinate time for lapse $N = 1/p_b \sqrt{p_c}$.
One finds that $p_b$ grows exponentially in this regime with respect to the proper time, whereas
$p_c$ attains a constant value. For the initial conditions in Fig. \ref{triads_bh}, we show that
such an asymptotic region occurs in positive time (see also Fig. \ref{_bh_asym}).
Note that in contrast to the Kantowski-Sachs spacetime with a negative cosmological constant, the transitions are not as
smooth which can be seen in the zoomed version of $p_b$ and $p_c$ in the above figure.
In the negative time, $p_b$ decreases whereas
$p_c$ increases after several cycles of bounces and recollapse as is shown in Fig. \ref{triads_bh}. For large positive time, the mean volume of the spacetime increases
in each cycle of loop quantum bounce and the classical recollapse. Finally, we note that in the asymptotic regime where $p_c$ attains a constant
value, holonomy corrections are significant similar to the case of asymptotic spacetime in the loop quantum model of Kantowski-Sachs spacetime shown in Fig. \ref{cos1fig}.
As in the case of the Kantowski-Sachs spacetime, this regime is not a solution of the classical Hamilton's equations for the Bianchi-III LRS
model with a negative cosmological constant. In the following section we discuss further properties of this asymptotic regime.

\section{Properties of the asymptotic spacetime with a constant $p_c$}
The main conclusion from the numerical simulations of the Kantowski-Sachs spacetimes with a positive and a
 negative cosmological constant, and the Bianchi-III LRS spacetime (or the higher genus black hole spacetime) with a negative cosmological
constant is that the
spacetime in one side of the temporal evolution after the singularity resolution has a constant value of triad $p_c$.
Let us look at some of the key features of this asymptotic regime. As was illustrated in
Fig. \ref{cos1fig}, in this regime $\cos(c \delta_c) = 0$. Thus, $c \delta_c$ takes a constant value, and
eq.(\ref{pc-eff}) implies that $\dot{p_c}=0$. Further, using eq.(\ref{b-eff}) we find that $b$ is a
constant in proper time for this asymptotic regime in Kantowski-Sachs model as well as the higher genus black hole interior. Since
$\delta_b = \sqrt{\Delta/p_c}$, in this regime $b \delta_b$ is also a constant.
Using the constancy of $b\delta_b$ and $c\delta_c$ in \eqref{pb-eff} one finds that  $\frac{\dot{p_b}}{p_b}$ is a constant. 
That is, $p_b$ expands exponentially in proper time, both for the Kantowski-Sachs and higher genus
black hole spacetimes. For the case of the higher genus black hole interior, where numerical simulations were carried with lapse $N=1/p_b \sqrt{p_c}$, $p_b$ is a constant with respec to the coordinate time in the asymptotic regime. 
Note that since $\delta_c = \sqrt{\Delta} p_c^{1/2}/p_b$, for $c\delta_c$ to be a constant,
$\frac{c}{p_b}$ has to be a constant. Hence $c$  also exhibits an exponential behavior with respect to proper time.
In the simulations discussed in Sec. III,  one finds that both $c$ and $p_b$ expand exponentially in such a way that $\frac{c}{p_b}$
is a constant in the asymptotic regime.

Since the triads are related to the metric components via equation \eqref{relationtometric}, knowing the asymptotic behavior of the triads allows  us to find the asymptotic behavior of the metric components.
The radius of the 2-sphere part, $g_{\theta\theta}=p_c$,  hence has the same asymptotic value as the triad $p_c$. Setting this asymptotic value to be $R_0^2$, we have asymptotically
\begin{equation}\label{asymptotic_gthetatheta}
 g_{\theta\theta}(\tau)=R_0^2,
\end{equation}
a constant value, where $\tau$ is the proper time. Also, it is obvious that in the asymptotic region, $g_{\phi\phi}=R_0^2\sin^2{\theta}$ for the Kantowski-Sachs spacetime and $g_{\phi\phi}=R_0^2\sinh^2{\theta}$ for the Bianchi III LRS spacetime. 

In the asymptotic regime, while the triad $p_c$ is constant, triad $p_b$ is expanding exponentially. Let us first consider the case when one reaches this asymptotic region in the forward evolution. Then, asymptotically $p_b=p_b^{(0)}e^{\alpha \tau}$,
where $\alpha$ is a positive constant. The coefficient $p_b^{(0)}$ is a positive constant which formally has the following meaning. If the exponential expansion began at some proper time $\tau_0$,
with the initial value of $p_b$ as $p_b^{(i)}$, then, $p_b^{(0)}=\frac{p_b^{(i)}}{e^{\alpha \tau_0}}$. Note that this is not the initial value of $p_b$ where the numerical evolution was started with at $\tau=0$.
Since $p_b$ scales linearly with the fiducial length, $p_b^{(0)}$ is also a fiducial cell dependent quantity. If the asymptotic region with the exponential behavior of $p_b$ had been in the backward evolution,
then $\alpha$ would have been a negative quantity and $p_b$ would increase exponentially as $\tau \rightarrow -\infty$. Either way,
\begin{equation}\label{asymptotic_gxx}
 g_{xx}(\tau)=\frac{p_b^2}{L_0^2p_c}=\frac{(p_b^{(0)})^2e^{2\alpha \tau}}{L_0^2 R_0^2}.
\end{equation}
Note that since $p_b^{(0)}$ depends on the fiducial length linearly, $g_{xx}$ is independent of the fiducial cell. Once we have the metric components from \eqref{asymptotic_gthetatheta} and \eqref{asymptotic_gxx}, we can analyze the properties of these asymptotic spacetimes
and ask if they satisfy the Einstein equations for some $T_{\mu\nu}$. We do this separately for the Kantowski-Sachs spacetime and the Bianchi III LRS spacetime in the following subsections.


We can now obtain the effective metric of the asymptotic spacetime for the Kantowski-Sachs case with a positive and a negative cosmological constant, and the higher genus black hole interior with a negative cosmological constant. 
Using equations \eqref{asymptotic_gthetatheta} and \eqref{asymptotic_gxx}, the line element of the emergent asymptotic spacetime in the LQC evolution of Kantowski-Sachs spacetime can be written as
\begin{equation}
 ds^2=-d\tau^2+\frac{(p_b^{(0)})^2e^{2\alpha \tau}}{L_0^2 R_0^2}dx^2+R_0^2\left(d\theta^2+\sin^2\theta d\phi^2\right).
\end{equation}
Since $\tau \rightarrow \rm \infty$ in the asymptotic region, it is permitted to substitute $e^{\alpha \tau}$ with $\cosh{(\alpha \tau)}$. Additionally, using the redefinition $x \rightarrow \frac{p_b^0}{L_0R_0}x$,
the line element can be written as
\begin{equation}\label{metric_chargednariai}
 ds^2=-d\tau^2+\cosh^2{(\alpha \tau)}dx^2+R_0^2\left(d\theta^2+\sin^2\theta d\phi^2\right).
\end{equation}
The above line element \eqref{metric_chargednariai} is formally similar to the Nariai metric written in its homogeneous form \cite{n1,Bousso,bv},
with the difference that for the Nariai metric $\alpha=\frac{1}{R_0}$. This formal similarity had led the earlier authors \cite{bv} to call the asymptotic spacetime obtained as Nariai-type spacetime. In classical GR, for $\alpha \neq 1/R_0$, the spacetime metric corresponds to a 
charged Nariai spacetime with a uniform electromagnetic field. Thus, the effective metric obtained above formally corresponds to a charged 
Nariai solution of the Einstein's theory. 
Such spacetimes are often
discussed in literature \cite{dias,cardoso,bousso2} in static coordinates. One can go to those coordinates using the transformations
 $k\tau \rightarrow \sinh(k\tau)$, $R=i\tau$ and $T=ix$. The line element 
\eqref{metric_chargednariai} becomes
\be
\label{metric3}
ds^2=-(1-k^2R^2)dT^2+\frac{dR^2}{1-k^2 R^2}+R_0^2 \left( d\theta^2 +\sin^2\theta d \phi^2 \right).
\ee
One can do yet another coordinate transformation $k^2=\frac{k_0}{R_0^2}$ and $\sin^2\chi=1-k^2R^2$, to write the static line element
in the form used in the references \cite{dias,cardoso,bousso2},
\be \label{metric}
ds^2=\frac{R_0^2}{k_0}\left(-\sin^2(\chi) d\tau^2 + d\chi^2 \right)+R_0^2 \left( d\theta^2 +\sin^2(\theta)d\phi^2 \right).
\ee
Thus we see that the asymptotic spacetime we obtain in the evolution of Kantowski-Sachs spacetime corresponds to a classical charged Nariai spacetime. However, the quantum spacetime is different
from these classical charged spacetimes in the sense that the `charge' in LQC evolution of Kantowski-Sachs spacetime is purely of quantum geometric origin and is not electromagnetic.

Similarly, one can write the line element for the asymptotic constant $p_c$ spacetime obtained in the LQC evolution of Bianchi III LRS spacetime (or higher genus black hole interior) with negative cosmological constant
using  \eqref{asymptotic_gthetatheta} and \eqref{asymptotic_gxx} together with $e^{\alpha \tau} \rightarrow \cosh{(\alpha \tau)}$ and $x \rightarrow \frac{p_b^0}{L_0R_0}x$ as,
\begin{equation}\label{metric_BR2}
 ds^2=-d\tau^2+\cosh^2{(\alpha \tau)}dx^2+R_0^2\left(d\theta^2+\sinh^2\theta d\phi^2\right).
\end{equation}

Following the coordinate transformations used to reach from \eqref{metric_chargednariai} to \eqref{metric}, one can write the line element \eqref{metric_BR2} in the static coordinates as
\be \label{metric_BR2_static}
ds^2=\frac{R_0^2}{k_0}\left(-\sin^2(\chi) d\tau^2 + d\chi^2 \right)+R_0^2 \left( d\theta^2 +\sinh^2\theta d\phi^2 \right).
\ee
This metric corresponds the anti Bertotti Robinson spacetime, which in classical GR is an electrovacuum solution. As in the case of emergent 'charged' Nariai spacetime discussed above, even though classically these spacetimes are solutions of Einstein equations with
matter as uniform electromagnetic field in the loop quantum case the energy momentum tensor in these emergent spacetime originates from the 
quantum geometry and is not electromagnetic in origin.

`Charged' Nariai and anti-Bertotti-Robinson spacetimes that emerge in LQC evolution of black hole interiors are product of constant curvature spaces.
`Charged' Nariai spacetime has a topology of $dS^2 \times \mathbb{S}^2$ where as the anti Bertotti-Robinson spacetime
has the $dS^2 \times \mathbb{H}^2$ topology.
Thus, both are product manifolds with each of the two manifolds ($t-x$ manifold whose curvature may be denoted by $k_+$ and
$\theta-\phi$ manifold whose curvature may be denoted by $k_-$) having constant curvature.
For a discussion of these product of constant curvature manifolds in classical GR we refer the reader to \cite{Dadhich}.\footnote{Nariai and Bertotti-Robinson spacetimes have also been shown to be solutions of modified Einstienian gravity in presence of one loop quantum corrections to the stress energy tensor, see for eg. Ref.\cite{sahni}.} For classical charged Nariai spacetime (as well as for the quantum ones with the non electric or magnetic charge),
both the $t-x$ manifold and $\theta-\phi$ manifold have positive curvatures. The special case of both these curvatures being equal $k_+=k_- > 0$ corresponds to the `uncharged' Nariai spacetime
(or just Nariai spacetime as it is commonly referred to in the literature). For anti-Bertotti-Robinson spacetime, $k_+$ is positive where as $k_-$ is negative.
The special case of $k_+=-k_-$ corresponds to a spacetime with no cosmological constant and is a conformally flat spacetime.
There are two other spacetimes (anti-Nariai and Bertotti-Robinson spacetimes) for different choices of the sign of $k_+$ and $k_-$ as tabulated below.

\begin{table}[!ht]
\centering
\begin{tabular}{|c c c  c c|}
\hline
\hspace{.5in}Type\hspace{.5in} &\hspace{0.5in}Topology\hspace{0.5in} & \hspace{.5in}$R^{t}_{t}$\hspace{.5in}  & \hspace{.5in}$R^2_2$\hspace{.5in} & \hspace{0.5in}$\tilde\Lambda$\hspace{0.5in} \\
\hline
Nariai & $dS_2 \times S^2$ & +ve & +ve & +ve \\

Anti-Nariai & $AdS_2 \times H^2$ & -ve & -ve & -ve\\

Bertotti-Robinson & $AdS_2 \times S^2$ &-ve & +ve & any sign\\

Anti-Bertotti-Robinson & $dS_2 \times H^2$ & +ve & -ve & any sign\\
\hline
\end{tabular}
\caption{Some features of (anti) Nariai and (anti) Bertotti-Robinson spacetimes in relation to the allowed signs of curvature and cosmological constant are shown.}
\label{table}
\end{table}

In the present context, the necessary condition to respect positivity of energy is $k_- > k_+$. 
This is never true in anti-Bertotti-Robinson spacetimes as $k_- < 0$ and $k_+ > 0$, and hence
they always violate positive energy conditions. Even though negative energy solutions of Einstein equations are deemed unphysical,
it is plausible that in scenarios where quantum geometric effects are important, the energy density may be allowed to be negative. Such a scenario where a modified gravity theory
gives rise to negative energy density has been observed earlier in the context of brane world models \cite{naresh_plb,naresh_neg_energy}.

\section{Emergent `charge' and cosmological constant in loop quantum cosmology}

The charged Nariai and anti Bertotti-Robinson solutions in classical general relativity have electromagnetic field as its matter content. The asymptotic spacetimes obtained in the loop quantum evolution of
Kantowski-Sachs and Bianchi III LRS spacetimes are identical to these classical solutions as far as the metric and curvature components are concerned. Hence from a computational point of view
determining the energy momentum tensor of these emergent spacetimes follow that of the corresponding general relativistic spacetime. However one must be cautious in interpreting the nature of
this energy momentum tensor.  Specifically, one should not attribute the energy momentum tensor to the existence of a uniform electromagnetic field as one would do in the classical
geometrodynamical setting. The `field' and `charge' arising in loop quantum evolution of black hole interiors is not electromagnetic but quantum geometric.
It is plausible that these quantum spacetimes are sourced by some energy momentum tensor of quantum gravitational origin but similar to that of the uniform electromagnetic field in its effect on curvature.
This is the sense in which we use the word `charge' or `field' to describe the matter content of emergent spacetimes. The situation is similar to the case of the bulk/brane models, where the Weyl curvature of the bulk induces a `charge' on the brane \cite{naresh_plb}.  
Further studies are required to exactly pin point the nature of this `charge' and
to determine if this charge is identical to the electromagnetic charge in geometrodynamics.
For now, we adopt the techniques of geometrodynamics \cite{misner} for computational purposes, relegating the interpretation of the `charge' to future works. In the following subsection we summarize the
calculation of energy momentum tensor and charge in a classical setting, which then will extended to the calculation of non electric (or magnetic) charge in the quantum spacetimes obtained in the evolution of
black hole interiors.

Let us first consider the classical aspects of charged Nariai and anti-Bertotti-Robinson spacetimes. 
In classical geometrodynamics, with a vanishing cosmological constant, for a spacetime to admit a uniform electromagnetic field,  the Ricci tensor components must satisfy $R^0_0 = R^1_1 = - R^2_2 = - R^3_3$. For a spacetime satisfying above conditions, the electromagnetic field tensor that satisfies the Einstein equations
\begin{equation}
 R^i_j -\frac{R}{2} g^i_j =\frac{1}{4\pi} \left(F^{ik}F_{jk}-\frac{F^{kl}F_{kl}}{4}g^i_j \right)
\end{equation}
can be found uniquely from the Ricci tensor \cite{misner}. This result was extended to the case in the presence of a non-zero cosmological constant by Bertotti \cite{bertotti}.

 The electromagnetic field tensor in charged Nariai spacetime \eqref{metric} is \cite{bousso2,dias}
 \be \label{fthetaphi}
F= q\sin{\theta} d\theta \wedge d\phi
\ee
when the field is purely magnetic and
\be
F=\frac{ q}{k_0}\sin{\chi} d\tau \wedge d\chi
\ee
when field is purely electric. Here $ q$ is the electric or magnetic charge. In geometrodynamics,  $ q$ is purely geometric in origin and has nothing to do with the quantized charge, of say an electron \cite{misner}.
The electromagnetic field is constant in spacetime and hence one might expect that there is no localized charge. This is indeed true, and one can see that the 4-current vanishes as
$F^{ik};_k=0$.
 Thus the field can be visualized as created by effective charges at the boundary
of the spacetime \cite{Dadhich}.
The energy momentum of the field is \cite{dias}
\be \label{tij}
\text{diag}(T^i_j)=(\frac{- q^2}{8 \pi R_0^4},\frac{- q^2}{8 \pi R_0^4},\frac{ q^2}{8 \pi R_0^4},\frac{ q^2}{8 \pi R_0^4}).
\ee
Since the above energy momentum tensor is traceless, the cosmological constant in the charged Nariai spacetime 
is determined by the Ricci scalar. The energy momentum tensor and cosmological constant of anti-Bertotti-Robinson spacetime can also be found in the same way.  The only difference for anti Bertotti-Robinson spacetime from the above calculation is that one should use $\sinh \theta$ instead
of $\sin \theta$ in \eqref{fthetaphi} and $-q^2$ instead of $q^2$ in \eqref{tij}.

The emergent loop quantum spacetimes obtained in Sec. III have the same curvature and Einstein tensors as the one discussed in the classical theory above.  The difference however is that for these quantum spacetimes the energy momentum tensor in  \eqref{tij} is purely of quantum geometric origin. Thus the electric
or magnetic charge $q$ in  \eqref{tij} should be substituted with an emergent `charge' $\tilde q$ which is neither electric nor magnetic. Similarly the cosmological constant is also of quantum geometric origin
and we refer to it as $\tilde \Lambda$. The emergent `charge' $\tilde q$ and $\tilde \Lambda$ is related to $k_0$ and $R_0$ in \eqref{metric} as in the classical case \cite{dias}
\be
\tilde q^2=\frac{k - k_0}{2}R_0^2=\frac{k -\alpha^2R_0^2}{2}R_0^2.
\ee
and 
\be
\tilde\Lambda=\frac{k + k_0}{2R_0^2}=\frac{k + \alpha^2R_0^2}{2R_0^2}.
\ee
where $k = +1$ for the `charged' Nariai spacetime and $k = -1$ for the anti-Bertotti-Robinson spacetime.  In the latter case, since the 
electric energy density is negative, the corresponding emergent charge satisfies $\tilde q^2<0$. Note that this interpretation is based on the appropriate choice of splitting of the constant curvatures $k_+$ and $k_-$ giving rise to a mixture of Nariai and (anti) Bertotti-Robinson spacetimes.

Using the asymptotic values of $\alpha^2$ and $R_0^2$ obtained numerically from the LQC evolution in Kantowski-Sachs and the higher genus black hole interior, the values of emergent `charge' $\tilde q$ and emergent cosmological constant $\tilde \Lambda$ can be computed. For the 
Kantowski-Sachs case, with $\Lambda$ in eq.(\ref{effham}) zero, the the emergent `charge' and cosmological constant of
the `charged' Nariai spacetime in geometrized units evaluates to $\tilde q^2=0.151$ and $ \tilde\Lambda=1.610$. 
The emergent `charge' and cosmological constant change by less than 0.1\% of these vacuum values for small magnitudes of cosmological constant ($|\rho_{\Lambda}| < 10^{-6}$). For higher values of $|\rho_{\Lambda}|$, the change in these emergent quantities from the values in the vacuum case are more appreciable. It was found that both the emergent `charge' and the
cosmological constant varies monotonically with $\rho_{\Lambda}$, such that  $\tilde q^2$ decreases and $\tilde \Lambda$ increases with increasing $\rho_{\Lambda}$.
For example, for $\rho_{\Lambda}=-10^{-2}$, we find $\tilde q^2 = 0.164$ and  $\tilde \Lambda = 1.487$ in geometrized units.
The same quantities turn out to be $0.139$ and $1.734$ respectively for $\rho_{\Lambda}=10^{-2}$.
Further work is required to find the precise dependence of these quantities on $\rho_{\Lambda}$. Also note that these quantities depend on the choice of $\Delta$ as
any change in the holonomy edge length will result in a change in asymptotic values of $p_c$ and $\frac{\dot{p}_b}{p_b}$.

We have already seen that the electric energy density developed in the anti-Bertotti-Robinson spacetime obtained in the evolution of Bianchi III LRS/higher genus black hole interior  is negative.  It turns out that the emergent
cosmological constant is also negative for these spacetimes. Once again, the values of these emergent quantities depend on $\rho_{\Lambda}$.
Unlike the case of Kantowski-Sachs spacetime, constant $p_c$ asymptotic regime occurs only if the cosmological constant is negative for the Bianchi III LRS spacetime in LQC evolution.
In this case, $\tilde q^2$ decreases with increasing (decreasing magnitude) $\rho_{\Lambda}$ whereas $\tilde \Lambda$ increases. For example,
for $\rho_{\Lambda}=-1 \times 10^{-8}$, we get $\tilde q^2=-0.050$ and $\tilde \Lambda = -4.997$ in geometrized units, whereas for
$\rho_{\Lambda}=-1 \times 10^{-2}$,  $\tilde q^2=-0.053$ and $\tilde \Lambda = -4.647$. The dependence of both the emergent quantities were seen to be approximately
linear with $\rho_{\Lambda}$, as in the case for emergent cosmological constant in LQC evolution of Kantowski-Sachs model.

In summary, the emergent asymptotic spacetimes in the loop quantum evolution of Kantowski-Sachs spacetime and Bianchi III LRS spacetime are a product of constant curvature spaces whose matter content is parameterized by
an emergent `charge' and cosmological constant, both of which are of quantum gravitational origin. The emergent charge in these spacetimes should not be apriori identified with electric or magnetic charge but is a result of
quantum geometry. This is similar to the emergence of a `tidal charge' in the brane world scenario discussed in \cite{naresh_plb} arising from the projection of the
Weyl tensor of a 5 dimensional brane world model on to a four dimensional brane.

\section{`Uncharged' Nariai spacetime and $\tilde \Lambda = 0$ anti-Bertotti-Robinson spacetimes}

A natural question to be posed in the light of last section is if loop quantum evolution of Kantowski-Sachs spacetime can lead to `uncharged' Nariai spacetime which corresponds to the special case of $k_+=k_-$
i.e, a spacetime where all the diagonal components of $R^i_j$ are equal.
Similarly one could also ask if the special anti Bertotti-Robinson solution with $k_+=-k_-$ (which corresponds to $\tilde \Lambda =0 $) may emerge from LQC evolution of Bianchi III LRS spacetime/higher genus black hole interior.

Let us rewrite the loop quantum Hamiltonian constraint (\ref{effham}) for the Kantowski-Sachs spacetime and for the Bianchi III LRS/higher genus black hole interior spacetime as
\be
\sin^2(b\delta_b)+2\sin(b\delta_b)\sin(c\delta_c)+\beta = 0
\ee
where $\beta=\gamma^2\Delta(k/p_c-\Lambda)$ with $k=1$ for Kantowski-Sachs spacetime, and  $k=-1$ for the higher genus black hole. Any solution with $\beta < -3$ is unphysical as it does not solve the Hamiltonian constraint.

We now check what additional conditions should be satisfied for a constant $p_c$ spacetime to exist. Since $\cos{(c\delta_c)}=0$ for derivative of $p_c$ to vanish,
$\sin{(c\delta_c)=\pm 1}$ in the constant $p_c$ regime. Hence, the Hamiltonian constraint for the constant $p_c$ regime is a quadratic equation in $\sin(b \delta_b)$,
\be
\sin^2(b\delta_b) \pm 2\sin(b\delta_b)+\beta = 0
\ee

Existence of a real solution requires $\beta \leq 1$. Hence $\beta$ is constrained to lie between -3 and 1. It is convenient to introduce 
$x = \sqrt{1 - \beta}$ which varies from 0 to 2, in terms of which $\alpha^2$ and $1/p_c$ become:
\begin{equation}
 \alpha^2=\frac{x^2}{\gamma^2\Delta}(2x-x^2), ~~\mathrm{and}~~\frac{1}{p_c}=\Lambda+\frac{k(1-x^2)}{\gamma^2 \Delta}.
\end{equation}

For the curvature components to satisfy $k_+ = k_-$ or $k_+ = - k_-$, or equivalently $\alpha^2=\pm\frac{1}{p_c}$, the following equation needs to be 
satisfied 
\be \label{fx}
x^4-2x^3+k(-x^2+1+\gamma^2 \Delta \Lambda) = 0 ~.
\ee
The function $f(x)=x^4-2x^3-kx^2$ with $x \in [0,2]$ has a range $[-4,0]$ for $k=1$ and a range $[0,4]$ for $k=-1$. Hence, the plausible range of cosmological constant is
\be \label{lambda_range}
\frac{-1}{\gamma^2 \Delta} \leq \Lambda \leq \frac{3}{\gamma^2 \Delta}
\ee
for the existence of both `uncharged' Nariai spacetime as well as anti Bertotti-Robinson spacetime with $\tilde \Lambda = 0$. So, for a $\Lambda$ lying in the above range, one may find a suitable value of $x$ which will yield a spacetime with equal magnitudes for all the non-vanishing Ricci components. 

Another condition which has to be satisfied for any constant $p_c$ spacetime is that $c \delta_c$ should be a constant. This is satisfied when $c/p_b$ is a constant,
which yields (using \eqref{pb-eff} and \eqref{c_eff})
\be \label{condition2}
1-x+\cos(\frac{\gamma^2 \Delta \Lambda +1 -x^2}{x \sqrt{2x-x^2}})=0 .
\ee
Since we need to solve equations \eqref{fx} and \eqref{condition2} simultaneously, we obtain 
\be \label{condition3}
1-x+\cos(x \sqrt{2x-x^2})=0 .
\ee
One of the solutions of the above equation is at $x=2$, which corresponds to $\alpha=0$ and such a spacetime will have constant $p_b$, not an exponentially expanding $p_b$. In fact such a spacetime will have all the
triads and cotriads constant. Since we are interested in obtaining an `uncharged' Nariai spacetime or anti-Bertotti-Robinson spacetime with a  vanishing cosmological constant, this solution will not be considered.
The second  root of \eqref{condition3} was numerically found to be $x=1.31646$. Note that this solution is true for both the Kantowski-Sachs and the higher genus black hole interior spacetimes. Knowing $x$, one can find the
asymptotic values of $p_c$ and $\alpha^2$ as $p_c=0.18697$ and $\alpha^2=5.34842$ in Planck units.

For the above values of asymptotic $p_c$ and $\alpha$, one can obtain an `uncharged' Nariai solution in the LQC evolution of Kantowski-Sachs spacetime or anti Bertotti-Robinson spacetime in the LQC evolution
of Bianchi III LRS spacetime.
The cosmological constant needed for obtaining such a Nariai spacetime turns out to be 7.86251 in Planck units. This is a huge value, especially since the corresponding energy density is around three fourth of the
critical energy density of isotropic loop quantum cosmology. For Bianchi III LRS spacetime, the cosmological constant needed to obtain an anti-Bertotti-Robinson spacetime without emergent cosmological
constant turned out to be -2.83432 in Planck units.

For these special spacetimes to emerge naturally from the loop quantum evolution of black hole interiors, the corresponding constant $p_c$ regime should be stable, which turns out to be not the case \cite{js3}. The stability was tested by choosing initial conditions from the exact solution, such that the Hamiltonian constraint is satisfied, and then using LQC equations of motion  to evolve the spacetime.
It was found that the `uncharged' Nariai spacetime evolves to a deSitter spacetime in both past and future evolution and thus is not stable.
This does not come as surprise given the high value of positive cosmological constant, for which generic initial conditions lead to de Sitter spacetime in both the forward and backward evolution of Kantowski-Sachs
universe in LQC.
The anti-Bertotti-Robinson solution with vanishing emergent cosmological constant was also found to be unstable. It evolved in to an anti-Bertotti-Robinson solution with a smaller constant $p_c$ (and thus a nonzero emergent cosmological constant) on one side
where as $p_c$ kept on increasing after each recollapse on the other side. In summary we did not find a stable emergent `uncharged' Nariai spacetime or anti-Bertotti-Robinson spacetime with vanishing emergent cosmological constant
in the loop quantum evolution of Kantowski-Sachs or Bianchi III LRS spacetimes. Thus whenever a constant $p_c$ regime emerges from the quantum evolution of these spacetimes, there is an associated emergent `charge'
and an emergent cosmological constant. In the presented scenario spacetime which is a  product of constant curvature spaces with equal magnitudes for all the non-vanishing Ricci tensor components seems to be disfavored in the LQC evolution of black hole interiors.

\section{Discussion}

Loop quantum evolution of Schwarzschild and higher genus black hole interiors were studied in a minisuperspace setting using their isometries to Kantowski-Sachs spacetime  (in \cite{bv}) and Bianchi III LRS
spacetime (in \cite{Brannlund}) respectively. These studies found that the black hole interior spacetime undergoes a quantum bounce and the classical central singularity is resolved, similar to the
singularity resolution in other loop quantum cosmological models. However, the post bounce evolution in these black hole interior models has the surprising feature that it asymptotes towards a spacetime
where the triad $p_c$ which is the radius of the two-sphere part attains a constant value where as the triad $p_b$ undergoes an exponential expansion. Also, these spacetimes were found to have non-negligible holonomy
corrections asymptotically after the bounce, i.e, the quantum geometric effects do not fade away after the bounce. Due to its similarity with the Nariai spacetime which also has a constant $g_{\theta \theta}$ and
an exponentially increasing $g_{xx}$, these asymptotic regions were termed `Nariai type' spacetimes in previous works \cite{bv,Brannlund}.
In this work, we re-examined the loop quantum evolution of black hole interior spacetimes in the presence of cosmological constant to study the asymptotic spacetime in detail assuming the validity of the effective spacetime description in LQC. 

We find that the asymptotic constant $p_c$ spacetime obtained in the effective loop quantum evolution of Kantowski-Sachs spacetime can be interpreted as a `charged' Nariai solution of classical GR.
The asymptotic solution obtained in the evolution of Bianchi III LRS spacetime with a negative cosmological constant (or the higher genus black hole interior) can be similarly interpreted as an anti-Bertotti-Robinson spacetime, again a solution to the Einstein equations.
Both these solutions are product spacetimes of constant curvature manifolds \cite{Dadhich} with $R^0_0=R^1_1= \pm  R^{2}_{2}=\pm R^{3}_{3}$. 
 The constancy of Ricci components was verified numerically for the asymptotic
spacetimes emergent in LQC evolution of black hole interiors. These curvatures were found to be Planckian even though the effective metric is a solution of the Einstein equations. Thus the emergent spacetime
has the peculiar nature of being isometric to a classical spacetime while the quantum gravity effects (via the holonomy corrections of LQC) are large. It is also noteworthy that classical `charged' Nariai
and anti-Bertotti-Robinson spacetimes are nonsingular and thus the geodesics in black hole interiors can be extended to infinity in the loop quantum evolution. Another striking feature of these spacetimes is the
existence of an emergent `charge' and cosmological constant - both of quantum geometric in origin. One could fine tune the asymptotic $p_c$ and $\frac{\dot{p}_b}{p_b}$ to obtain `uncharged' Nariai spacetime
or anti-Bertotti-Robinson spacetime with vanishing cosmological constant.  However these finetuned spacetimes turn out to be unstable and evolve to more generic `charged' Nariai spacetime or anti-Bertotti-Robinson spacetime with cosmological constant. Thus emergence of the quantum geometric `charge' and cosmological constant seems to be inevitable in the LQC evolution of black hole interior spacetimes. The basic property of emergent spacetime is that it is a product of two spaces of unequal constant curvatures. The interpretation of these constant curvature spaces in LQC as `charged' Nariai and anti-Bertotti-Robinson spacetimes comes about by a proper splitting of the two curvature constants of product spaces.
This is indeed remarkable that emergent spacetime after the bounce is classical GR solution.

The emergence of a quantum geometric `charge' in the LQC evolution of black hole interior spacetimes is in parallel to a similar result in the brane world scenario \cite{naresh_plb}. There, the projection of the
Weyl tensor of a 5 dimensional bulk spacetime on to a four dimensional brane gave rise to Weyl or `tidal' charge (and not an electric charge)  which was taken to be negative on physical considerations. The metric however had exactly the same form as that of  Reissner-Nordstrom solution with $Q^2 \rightarrow -Q^2$. The parallel between the current work and the findings of \cite{naresh_plb} are striking. In both cases, a theory of modified general relativity gives rise to an emergent `charge' which mimics an electric or magnetic charge in the way it affects the geometry of spacetime, but which has its origin not in electromagnetism but in the modifications to general relativity. Thus this emergent 'charge' is purely gravitational in nature. However it seems that we are finding evidence that attempts to capture quantum corrections to general relativity through such a `charge'. Much further work and analysis is required to further nail its actual character and significance.

\acknowledgements
We thank Patrick Das Gupta and Miguel Megevand for helpful discussions. This work is facilitated by ND's visit to Louisiana State University, and that of PS to Center for Theoretical Physics, Jamia Millia University. This work is supported by NSF
grants PHYS1068743 and PHYS1404240.

\end{document}